\documentclass[twocolumn]{aastex631}

\usepackage{pbox, cellspace}
\usepackage{hyperref}
\renewcommand{\arraystretch}{2}

\usepackage[T1]{fontenc}
\usepackage{graphicx}	
\usepackage{amsmath}	
\usepackage{amssymb}	
\usepackage{wrapfig}
\usepackage{natbib}
\usepackage{xcolor}
\usepackage{ulem}
\usepackage{soul}
\usepackage{array}
\usepackage{tabularx}
\definecolor{blazeorange}{rgb}{1.0, 0.4, 0.0}
\definecolor{seagreen}{rgb}{0.18, 0.55, 0.34}
\definecolor{rufous}{rgb}{0.66, 0.11, 0.03}
\definecolor{royalfuchsia}{rgb}{0.79, 0.17, 0.57}
\definecolor{scarlet}{rgb}{1.0, 0.13, 0.0}
\definecolor{royalpurple}{rgb}{0.47, 0.32, 0.66}
\definecolor{darkblue}{rgb}{0, 0, 0.66}

\DeclareRobustCommand{\VAN}[3]{#2}
\let\VANthebibliography\thebibliography
\def\thebibliography{\DeclareRobustCommand{\VAN}[3]{##3}\VANthebibliography}

\hypersetup{linkcolor=blue,citecolor=blue,filecolor=cyan,urlcolor=magenta}
\usepackage{newtxtext,newtxmath}
\usepackage{threeparttable}
\usepackage{mathtools}

\def\d{\mathrm{d}}

\newcommand{\lta}{\lower 2pt \hbox{$\, \buildrel {\scriptstyle <}\over {\scriptstyle \sim}\,$}}
\newcommand{\gta}{\lower 2pt \hbox{$\, \buildrel {\scriptstyle >}\over {\scriptstyle \sim}\,$}}

\begin{document}

\title{The kinetic-energy bottleneck in Fast Radio Burst models}

\correspondingauthor{Paz Beniamini}
\email{pazb@openu.ac.il}

\author[0000-0001-7833-1043]{Paz Beniamini}
\affiliation{Department of Natural Sciences, The Open University of Israel, P.O Box 808, Ra'anana 4353701, Israel}
\affiliation{Astrophysics Research Center of the Open university (ARCO), The Open University of Israel, P.O Box 808, Ra'anana 4353701, Israel}
\affiliation{Department of Physics, The George Washington University, 725 21st Street NW, Washington, DC 20052, USA}

\author{Pawan Kumar}
\affiliation{Department of Astronomy, University of Texas at Austin, Austin, TX 78712, USA}

\begin{abstract}
Most Fast Radio Burst (FRB) models invoke a two-step process in which energy released by the central engine is converted into particle kinetic energy and subsequently radiated as coherent GHz emission. We derive model-independent constraints on this picture and apply them to the three main classes of FRB models.
(i) Inner-magnetospheric models violate brightness-temperature and kinetic-luminosity limits unless particles are continuously re-accelerated in situ. Magnetar-strength fields can provide the required parallel electric field out to $R\lesssim10^{10}\mathrm{cm}$. Monster shocks provide continuous acceleration but, under a single-pass kinetic-energy interpretation, require extreme particle densities and predict emission well above GHz frequencies.
Even mechanisms that bypass the kinetic bottleneck through conservative wave conversion remain subject to charge availability, which independently limits magnetospheric GHz production to $\lesssim 10^{10}$\,cm for fiducial parameters.
(ii) Light-cylinder forced-reconnection naturally provides continuous acceleration, but typically converts only $\lesssim10^{-6}$ of the injected energy into radio emission.
(iii) External-shock maser models satisfy the kinetic-luminosity and brightness-temperature constraints, but the upstream wind is unavoidably optically thick to induced Compton scattering, independent of the model’s principal parameters. Proposed escape routes - emission above the maser peak or upstream magnetization $\sigma_{\rm w}\gtrsim30$ imply very low efficiencies, while the former also conflicts with narrow FRB spectra.
We conclude that near-surface magnetospheric models with continuous particle acceleration remain the most promising scenario, provided the radiation can escape the magnetosphere.
\end{abstract}

\keywords{fast radio bursts -- stars: neutron -- stars: magnetars -- radiation mechanisms: non-thermal -- shock waves}

\section{Introduction} 
\label{sec:intro}
A common requirement underlies almost all proposed mechanisms for Fast Radio Burst (FRB) emission: regardless of the specific microphysical process or the location of the emission site, the energy powering the burst must be carried in kinetic form, by a relativistic particle beam or bulk flow, at the point where coherent radiation is produced. This requirement emerges across three broad classes of models, distinguished primarily by their emission radius, and imposes stringent constraints on the generation, transport, and dissipation of kinetic energy in any viable FRB source.

Magnetospheric mechanisms for FRB emission divide broadly into categories distinguished by how coherence is achieved. In coherent-antenna scenarios  \citep{1971ApJ...164..529S,1975ApJ...196...51R,1976MNRAS.177..109B,1979SSRv...24..567C,Kumar+17,YangZhang2018,LuKumar2018b,Wadiasingh2019,KumarBosnjak2020,Cooper2021,Cooper2024}, particles are organized into bunches by an external driver - for example, charge-starvation gaps, Alfv\'en waves, magnetic reconnection structures, or current sheets - and emit coherent curvature or related radiation as they propagate along curved field lines. In plasma maser scenarios \citep{Lyutikov2021}, coherent emission arises through beam-plasma instabilities such as Cherenkov-drift, cyclotron-Cherenkov, and related mechanisms, originally developed in the contexts of laser-plasma physics and coherent radio emission from ordinary pulsars. A third subclass invokes inverse Compton scattering of low-frequency magnetospheric waves by charge bunches \citep{Zhang2022}, with the scattering process replacing the direct radiation mechanism. Despite differences in microphysics, all three classes share a common architecture: a relativistic particle beam is launched at some inner location, propagates to a distinct emission site, and converts a fraction of its kinetic energy into coherent radio. In each case, the achievable radio luminosity scales with the kinetic energy density of the beam at the emission site.

In intermediate scenarios, such as forced magnetic reconnection near the light cylinder \citep{Lyubarsky2020,Mahlmann2022}, magnetic free energy serves as the ultimate reservoir but is first converted into an MHD wave which escapes to outside the light-cylinder where it triggers reconnection. A fraction of the dissipated energy released by reconnection can accelerate particles and then be converted to radio waves.
Similarly, in far-away blast-wave models, such as synchrotron maser emission at ultra-relativistic magnetized shocks \citep{Lyubarsky14,Beloborodov17,Beloborodov19,Metzger+19,Margalit+20}, the burst energy is transported outward as the bulk kinetic energy of a relativistic shell and a small fraction of it is converted into coherent radio emission downstream of the shock. In this case the observed luminosity directly reflects the instantaneous kinetic energy flux of the ejecta at the emission radius.

While plasma beam models were primarily formulated and validated in the pulsar context, their application to FRBs represents a major, and qualitative, departure. Cosmological FRBs reach isotropic equivalent luminosities of $L_{\rm FRB}\approx 10^{41}-10^{43}\mbox{erg s}^{-1}$, exceeding those of radio pulsars by $\sim 12$ orders of magnitude in luminosity and brightness temperature. This enormous gap means that beam particles must lose energy on extremely short timescales, and that the magnetosphere must supply and sustain energy fluxes with no pulsar analogue. 
In other words, even if mechanisms known from pulsar physics can in principle operate, we must first verify that the FRB source can generate and deliver the required kinetic power to the emission site. 

The constraints derived in this paper are complementary to a growing body of observational work that has sought to localize the FRB emission region. Frequency-time downward drifting (the ``sad trombone" effect) observed in many repeaters has been interpreted as radius-to-frequency mapping along curved magnetic field lines (e.g. \citealt{Hessels+19}), suggesting an origin in a strongly magnetized environment close to the central engine. Strong temporal variability on microseconds or less \citep{Majid2021,Nimmo2021,Snelders2023,Hewitt2023} and sharp spectral features down to a few MHz in width ($\Delta\nu/\nu \ll 1$; \citealt{Shannon2018,Zhu2025}) imply a compact emitting region  \citep{BK2020,Kumar2024}. Scintillation analysis has placed direct upper limits on the transverse source size of $\lesssim 3\times 10^{4}$\,km \citep{KBGC2024,Nimmo2025}, constraining models in which radio emission is produced far from the compact object. Rapid polarization-angle variations of bursts, including both smooth swings and abrupt jumps \citep{Niu2024,Mckinven2025,Manaswini2026}, have likewise been argued to reflect an ordered magnetic geometry associated with the central engine \citep{bk2025,2026ApJ...997...37Q}. Together, these observations suggest that the emission may originate in a compact region close to the central engine, possibly within or just outside the light cylinder.

At the same time, identifying the emission site does not uniquely determine the underlying emission mechanism. Even within the framework of near-engine models, considerable effort has been devoted to understanding whether coherent GHz radiation generated deep in a magnetar magnetosphere can escape. This is a major open question in FRB research which we do not attempt to resolve here. That said, due to the importance of this issue and its relevance for comparing the viability of nearby and far-away models, we include a dedicated discussion of the escape viability when addressing magnetospheric models in \S \ref{sec:magnetospheric}.

Motivated by these considerations, we ask here a more general question: what kinematic and energetic constraints must any coherent emission mechanism satisfy? We find that the energy bottleneck identified here arises generically across a broad class of models including nearby and far-away models and is largely independent of the detailed microphysics responsible for generating the observed radio emission.

\section{General kinetic bottleneck argument}
\label{sec:GeneralBottleneck}
Typical macroscopic disturbances that may trigger neutron-star activity (e.g. crustal oscillations, large reconnection events, etc.) are associated with timescales of order $10^{-3}\mbox{ s}\gg\nu_{\rm FRB}^{-1}$. Consequently, the macroscopic driver contains little power at GHz frequencies and a key component of any FRB model is to convert the appropriate power (typically from kHz) to GHz frequencies. As mentioned in \S \ref{sec:intro}, in many proposed models, this is done by having a significant energy channeled via a kinetic component. In \S \ref{sec:beam} we investigate the implications of this setup. We then turn in \S \ref{sec:applicable} to examine the realm of applicability of those implications. In \S \ref{sec:charge} we discuss the closely related issue of charge availability, leading to an independent constraint, applicable even beyond the kinetic bottleneck argument.

\subsection{Particle beam requirements}
\label{sec:beam}
Consider a cold particle beam with Lorentz factor (LF) $\gamma_j$, particle mass $m$, and number density $n$ measured in the neutron-star frame. Its isotropic-equivalent kinetic luminosity is
\begin{equation}
L_j = 4\pi R^2 m n \gamma_j c^3 .
\end{equation}

In the context of neutron-star magnetospheres, it is instructive to compare the densities to the Goldreich-Julian value \citep{GoldreichJulian1969}
\begin{equation}
 n_{\rm GJ} \approx {\Omega B_0\over e c \pi} \left({R_{\rm 0}\over R}\right)^3 \sim 2.2 \times 10^{14}
    {\rm \; cm}^{-3}\, B_{0,15}\Omega_1 \left({R_{\rm 0}\over R}\right)^3.
    \label{nGJ}
\end{equation}
where $R_{\rm 0}$, $\Omega$ and $B_0$ are the radius, rotation frequency and surface B-field strength of a neutron star, and $R$ is the distance to the center of the star.
We denote by $\xi=n/n_{\rm GJ}\geq 1$ the multiplicity factor.
For e$^+-e^-$ flows, $\xi$ grows through pair cascades and represents the number of pairs produced per primary particle. 
Theoretical studies of polar cap cascades in rotation-powered pulsars, incorporating curvature radiation, synchrotron radiation, and resonant inverse Compton scattering, establish a robust upper limit of $\xi\lesssim {\rm few} \times 10^5$ \citep{HardingMuslimov2011,TimokhinHarding2019}. Outer gap models yield lower values, $\xi\lesssim 10^4$ \citep{Hirotani2006}. For magnetar-strength fields, $B\gtrsim 10^{14}\,$G, additional QED processes, such as magnetic photon splitting, can suppress pair creation, although the degree of suppression depends on the still-debated question of which polarization modes participate in the splitting \citep{BaringHarding2001,Harding2025}.
Within the twisted magnetosphere framework \citep{BeloborodovThompson2007,Beloborodov2013}  pair avalanches are driven by electric currents flowing along closed field loops, sustained by crustal shearing of the magnetic field. The resulting charge densities can exceed the Goldreich-Julian value by factors of $\xi \sim 10^2-10^6$ depending on the twist angle and loop geometry. Recent Monte Carlo cascade simulations in this context, which self-consistently include one-photon pair production, photon splitting, and resonant inverse Compton scattering on closed magnetar field loops, find multiplicities of $\xi\sim 10-10^3$ per primary when radiative energy losses are accounted for, rising to $\xi \sim 10^4-10^5$ when such losses are neglected \citep{Harding2025}. Below we consider $\xi \lesssim 10^5$ as a conservative upper limit\footnote{The quoted estimates refer to steady or quasi-steady cascades, whereas FRBs may arise during strongly non-stationary episodes in which transient multiplicities could exceed these values. However, reconciling the particle-number (Eqs. \ref{eq:Nreq},\ref{eq:Nmag}) and brightness-temperature (Eq. \ref{Tb-max}) constraints with observations would require $\xi\gtrsim10^{12}$, seven orders of magnitude above the quasi-steady limit, sustained over the emitting region for the burst duration. Moreover, pairs produced in runaway discharges are born with a broad distribution of parallel momenta ($\Delta\gamma/\gamma\sim1$), which suppresses maser growth and disperses coherent bunches. Finally, the resulting density increase exacerbates the escape problem discussed in \S \ref{sec:monster}, \S \ref{sec:escape}. Therefore, temporary enhanced multiplicity does not naturally provide a viable way around the bottleneck.}.

In the beam rest frame, the causally connected length scale is $R/\gamma_j$, and the energy contained within this region is
$\delta\epsilon'_j \sim 4\pi R^3 m n c^2/\gamma_j^2$. This energy cannot be dissipated faster than the light-crossing time $R/(c\gamma_j)$, corresponding to an observer-frame duration
$\delta t_{\rm obs} \sim R/(c\gamma_j^2)$. Even assuming 100\% efficiency, the maximum observable luminosity is therefore $\gamma_j \delta\epsilon'_j/\delta t_{\rm obs}\sim 4\pi R^2 m n c^3 \gamma_j$, up to factors of order unity. This limit can be exceeded only if beam energy is temporarily stored in a region much smaller than the beam itself and radiated on its light-crossing time. However, plasma-maser models for FRBs typically assume that the beam excites waves that propagate outward and occupy an increasing volume, rather than being spatially compressed. Consequently, the luminosity estimated above is a generous upper bound that applies generically to beam-driven plasma models.

If a fraction $\varepsilon_{_{\rm rad}}$ of the observer frame beam power emerges as coherent GHz emission, the resulting radio luminosity is
\begin{eqnarray}
\label{eq:Lmax}
&L_{\rm rad}\! \sim \! 4\pi \varepsilon_{_{\rm rad}} R^2 m n c^3 \gamma_j \!\approx \!3\times 10^{36}\mbox{erg s}^{-1} \varepsilon_{_{\rm rad}} R_9^2 n_{10}\gamma_{j,3} \nonumber \\
&= 6.8\times 10^{36}\mbox{erg s}^{-1} \varepsilon_{_{\rm rad}} R_9^{-1} B_{0,15}\Omega_1\xi_5\gamma_{j,3}.
\end{eqnarray}
While this can easily accommodate the range of radio luminosities observed in pulsars $L_{\rm rad}\sim 10^{28}-10^{31}\mbox{erg s}^{-1}$ \citep{ATNF}, it falls many orders of magnitude below typical FRB luminosities ($L_{\rm FRB}\sim 10^{42}\mbox{erg s}^{-1}$ and up to $10^{45}\mbox{erg s}^{-1}$ in some cases), even for magnetospheric emission with $\xi=10^5$, from close to the neutron star surface and before accounting for the typically low efficiency of radio emission expected in FRB models (i.e. normalizing by $\varepsilon_{_{\rm rad}}=1$ above).

A second way of expressing the kinetic beam constraint is in terms of the brightness temperature. An observer detecting a burst of duration $t_{\rm FRB}$ would infer a brightness temperature
\begin{eqnarray}
& T_B \sim \frac{\varepsilon_{_{\rm rad}} R^2 m n \gamma_j c^3}
{2 k_B \nu^3 t_{\rm FRB}^2}
\sim 9\times 10^{29}\,{\rm K}\,
\frac{\varepsilon_{_{\rm rad}} R_9^2 n_{10} \gamma_{j,3}}
{\nu_9^3 t_{\rm FRB,-3}^2}\nonumber \\
& =2\times 10^{30}\,{\rm K}\,
\frac{\varepsilon_{_{\rm rad}} B_{0,15}\Omega_1 \xi_5 \gamma_{j,3}}
{\nu_9^3 t_{\rm FRB,-3}^2 R_9}.
  \label{Tb-max}
\end{eqnarray}
The source's relativistic motion toward the observer does not affect the inferred $T_B$,  since bulk motion does not enter into the observationally inferred $T_B$. Instead, the true brightness temperature in the compact-object rest frame and in the source comoving frame, however, depends strongly on $\gamma_j$.

Observed FRBs reach brightness temperatures as large as $T_B \sim 10^{37}\,\mathrm{K}$. Even adopting the optimistic value $\varepsilon_{_{\rm rad}} \sim 10^{-2}$, beam-powered magnetospheric emission falls short by more than seven orders of magnitude. 
Reproducing the observed $T_B$ would instead require $\gamma_j\sim 10^{10}$, which is unphysical. By contrast, beam-driven plasma masers readily account for radio pulsars, for which $T_{\rm B,obs} \sim 10^{25}$ K (however, Crab nano-shots reach $T_B$ comparable to FRBs - the constraints on their underlying emission are discussed further in \S \ref{sec:intermediate}), well within the bound implied by Eq.~\ref{Tb-max}. 

The analysis of $L_{\rm rad}$ and $T_{\rm B}$ already assumes that particles lose their kinetic energy on the dynamical timescale $R/(c\gamma_j)$ in the comoving frame, yet this proves insufficient by orders of magnitude. The only apparent remedy would be for particles to transfer their energy to plasma waves on a timescale shorter than the comoving dynamical time by a factor of $\sim 3\times 10^5$, while simultaneously being re-accelerated \emph{in situ} by a field-aligned electric field to prevent their LFs from decreasing. 

An independent and direct way to see the need for re-accelerating particles goes as follows. Considering the typical isotropic equivalent luminosity and duration of FRBs, we can estimate the isotropic equivalent energy that powered the burst $\epsilon_{\rm j,iso}\!\sim \!L_{\rm FRB}t_{\rm FRB}\varepsilon_{\rm rad}^{-1}\approx 10^{39}L_{\rm FRB,42}t_{\rm FRB,-3}\varepsilon_{\rm rad}^{-1}\mbox{ erg}$ where $\varepsilon_{\rm rad}$ is the radiative efficiency in the observer frame. At the same time, the energy per electron is $\gamma_j m c^2$. Thus, if there is no re-acceleration of electrons during the FRB emission, then the required isotropic equivalent electron number is
\begin{eqnarray}
\label{eq:Nreq}
   N_{\rm e,iso}^{\rm req}\!=\!
    \left\{ \begin{array}{ll}10^{42} L_{\rm FRB,42}t_{\rm FRB,-3}\gamma_{j,3}^{-1}\varepsilon_{\rm rad}^{-1} & \gamma_j\theta_j>1 \\
10^{42} L_{\rm FRB,42}t_{\rm FRB,-3}\gamma_{j,3}^{-1}\varepsilon_{\rm rad}^{-1} (\gamma_j\theta_j)^{-2} &  \gamma_j\theta_j<1
\end{array}\right. 
\end{eqnarray}
where $\theta_j$ is the opening angle of the beam, and the second line shows that for narrow beams and a fixed observed luminosity, the inferred $N_{\rm e,iso}^{\rm req}$ increases even further.
Typically, for realistic coherent emission mechanisms, we expect $\varepsilon_{\rm rad}\lesssim 10^{-3}$ and therefore $N_{\rm e,iso}^{\rm req}>10^{45}$.
This is much larger than the number of electrons within the entire magnetosphere of a NS, 
\begin{equation}
\label{eq:Nmag}
    N_{\rm mag} \approx \frac{4 \pi}{3} \xi n_{\rm GJ}(R_{\rm LC}) R_{\rm LC}^3\approx  10^{38} \xi_5 B_{0,15} \Omega_1\ll N_{\rm e,iso}^{\rm req}.
\end{equation}
This simple counting argument directly shows that for any plausible value of $\xi$, magnetospheric and intermediate FRB emission models must involve re-acceleration to reproduce the observed radiation. 

This shortcoming illustrated by the luminosity and brightness temperature constraints can also be expressed in terms of the cooling-time.
The reasoning, provided below, arrives at the same conclusion expressed differently, in a way which brings to the front the per-particle energetics and motivates the bunch-size estimates that follow.
Instead of asking how much energy is available within a causally connected region, we now ask how quickly an individual particle must radiate to sustain the observed luminosity. We find that the required timescale is so short as to be physically untenable.

The emission from a single particle within the beaming cone of opening angle $\gamma_j^{-1}$ is boosted in the observer frame according to $P_{\rm obs} = P'_{\rm em} \gamma_j^4$, where $P_{\rm em} = P'_{\rm em}$ is the Lorentz invariant emitted power. Allowing for coherent emission from bunches - where `bunch' denotes any localized region of particles whose emission adds in phase, sustained over the radiation timescale of interest rather than over the dynamical lifetime of the source - the isotropic-equivalent observed luminosity is $L_{\rm rad} = N^2 N_{\rm b} P_{\rm obs}$ where $N$ is the number of emitting electrons per bunch, and $N_{\rm b}$ is the number of bunches within the beaming cone. We note that objections raised in the pulsar context regarding the dynamical fragility of bunches \citep{1978ApJ...225..557M,2008AIPC..983...29L} are significantly altered with regards to FRB emission, which is a transient phenomenon: bunches need at most to persist over a single FRB duration, not over rotation periods.
The total isotropic equivalent number of electrons that are radiating at a fixed observer time is $N_{\rm tot,iso} \sim \gamma_j^2 N N_{\rm b}$.

In the central-engine frame, the cooling time is determined by the condition $N^2 P_{\rm em} t_{\rm c,CE} = N \gamma_j m c^2$ (where CE here stands for central engine). The observer frame time is shorter by a factor of $\gamma_j^2$, giving
\begin{eqnarray}
\label{eq:tcmax}
  &  t_{\rm c,obs} = \frac{m c^2}{\gamma_j N P_{\rm em}} = \frac{N_{\rm tot,iso} \gamma_j m c^2}{L_{\rm rad}} \lesssim \frac{2\pi n R^3 m c^2}{\gamma_j L_{\rm rad}} \\& = 5 \times 10^{-14}\, {\rm s}\, n_{10} R_9^3 \gamma_{j,3}^{-1} L_{\rm rad,42}^{-1},\nonumber \\ 
  & = 1.1 \times 10^{-13}\, {\rm s}\, B_{0,15}\Omega_1\xi_5 \gamma_{j,3}^{-1} L_{\rm rad,42}^{-1}\nonumber 
\end{eqnarray}
where we have taken the radial width of the emitting region to be the causally connected length $R/(2\gamma_j^2)$, so that $N_{\rm tot,iso} < 2\pi n R^3 / \gamma_j^2$. While it might appear counterintuitive that $t_{\rm c,obs}$ increases with $\xi$, this is merely a consequence of considering a given observed luminosity (recall that  $P_{\rm em}\propto L_{\rm rad}/ N^2$). The advantage of using $L_{\rm rad}$ is that it is directly observable, as opposed to the apriori unknown $P_{\rm em}$.

Since $t_{\rm c,obs} \ll t_{\rm FRB}\sim 1$\,ms, emission sustained over the observed burst duration cannot be maintained without a continuous supply of fresh kinetic energy to particles. The problem is in fact more severe: with such a short cooling time, and no continuous acceleration, the characteristic frequency of the emergent radiation is limited by the uncertainty principle.
The wavetrain duration $t_{\rm c,obs}\sim 10^{-13}\mbox{ s}$ implies a minimum bandwidth
$\Delta \nu\gtrsim t_{\rm c,obs}^{-1}\approx 10^{13}\mbox{ Hz}$, well above the radio band. The emission must therefore be broadband infrared rather than the narrow $\sim $GHz band actually observed.

The very short cooling time in Eq. \ref{eq:tcmax}, implies that only a small number of particles can become mutually coherent through self-organization (via radiation reaction or through plasma instabilities such as two-stream or modulational instabilities). In the beam rest frame, without continuous re-acceleration of the particles, the distance over which electrons can be causally connected and produce phase-coherent radiation in the observer's frame is $\ell'\sim c t_{c}'=ct_{\rm c,obs}\gamma_j\ll \gamma_j \lambda/(2\pi)$ where $\lambda\approx 30$\,cm is the wavelength of FRB radiation. In case particles are instead continuously accelerated and a balance is achieved between acceleration and radiation, then $\ell'\sim \gamma_j \lambda/(2\pi)$\footnote{The maximum possible size of $\ell'$ is the causally connected length $R/(2\gamma_j)$ which may be much larger than $\gamma_j \lambda$. However, special additional requirements must be put in place for this larger region to govern the maximum possible bunch size.}.
Multiplying the volume by $n'=n/\gamma$, we obtain the maximum number of electrons per bunch:
\begin{eqnarray}
\label{eq:Nscmax}
& N_{\rm sc,max}=n'\frac{4\pi}{3} \ell'^3 \lesssim\\
&  \left\{ \begin{array}{ll}3.7\times 10^{9} B_{0,15}^4\Omega_1^4 R_9^{-3}\xi_5^4\gamma_{j,3}^{-1}L_{\rm rad,42}^{-3} & \mbox{single shot acc.} \\
10^{19} B_{0,15}\Omega_1 R_9^{-3}\xi_5\gamma_{j,3}^{2}\nu_9^{-3} &  \mbox{continuous acc.}
\end{array}\right. \nonumber 
\end{eqnarray}
For single shot acceleration, the result is a relatively small number -- and as such it severely limits the gain in efficiency due to coherent emission. Note that the limit given by Eq. \ref{eq:Nscmax} only applies so long as the emission process relies on self-coordination of the emitting particles as a mechanism for producing coherence. If there is an external bunching agent (driver imposed coherence)---such as density modulations from two-stream instabilities or other plasma instabilities, charge-starvation gaps, or magnetic reconnection structures---the transverse size of the bunch can be much larger. In this case, it may extend up to the Fresnel length, \(R_{\perp} = (2R\lambda)^{1/2}\) \citep{Kumar+17}. However, the radial width remains limited by \(\ell'\).
This relies on there being a background field that is coherent over a length scale at least as large as $R_{\perp}$ and on the acceleration of electrons being governed entirely by this background field. It is then possible for a strong pulse to sweep through this region and simultaneously accelerate all the electrons, causing them to gyrate and radiate in phase at the observer position without requiring self-coordination\footnote{In order for the per-electron energy loss to scale as \(N P_{\rm em}\), rather than \(P_{\rm em}\), the bunch must radiate coherently rather than as individual particles. This requires the particles to establish dynamical coupling. In the comoving frame, the dynamical timescale is \(\sim R/(\gamma_j c) \sim 3\times 10^{-5}R_9\gamma_{\rm j,3}^{-1}\,\mathrm{s}\). For $R\gtrsim 7\times 10^7$\,cm, this is longer than \(R_{\perp}/c \approx 8 \times 10^{-6} R_9^{1/2}\nu_9^{-1/2}\,\mathrm{s}\), and causal contact across the bunch is established rapidly. The assumption that all particles in the bunch are dynamically coupled is justified in this case.}.
This leads to a larger maximum number of particles per bunch in the continuous case \footnote{For single shot acceleration, even with an external bunching mechanism, the lateral size is limited by $\min[R/\gamma, R_{\perp},ct_c']$. For typical FRB parameters, $ct'_c$ is much smaller than the other two length scales (see Eq. \ref{eq:tcmax}). We therefore use this scale to estimate $N_{\rm ex,max}$ in the top line of Eq. \ref{eq:Nexmax}.}
\begin{eqnarray}
\label{eq:Nexmax}
 &   N_{\rm ex,max}= \left\{ \begin{array}{ll} n'\frac{4\pi}{3} \ell'^3 & \mbox{single shot acc.} \\
n'\frac{8\pi}{3} R\lambda \ell' &  \mbox{continuous acc.}
\end{array}\right. \\
 &   \lesssim \left\{ \begin{array}{ll} 3.7\times 10^{9} B_{0,15}^4\Omega_1^4 R_9^{-3}\xi_5^4\gamma_{j,3}^{-1}L_{\rm rad,42}^{-3} & \mbox{single shot acc.} \\
2.6\times10^{22} B_{0,15}\Omega_1 R_9^{-2}\xi_5\nu_9^{-2} &  \mbox{continuous acc.}
\end{array}\right. \nonumber 
\end{eqnarray}
The bounds derived in this section are mechanism-independent in the following sense: regardless of whether coherence is achieved by an antenna mechanism, a plasma maser, or an inverse Compton scattering process, the radio output is limited by the kinetic energy density of the beam at the emission site. The brightness temperature, luminosity and cooling time bounds follow from this kinematic statement alone and do not depend on which microphysical channel converts kinetic energy into coherent radio waves.

\begin{figure}[ht]
\centering
\includegraphics[width=0.5\textwidth]{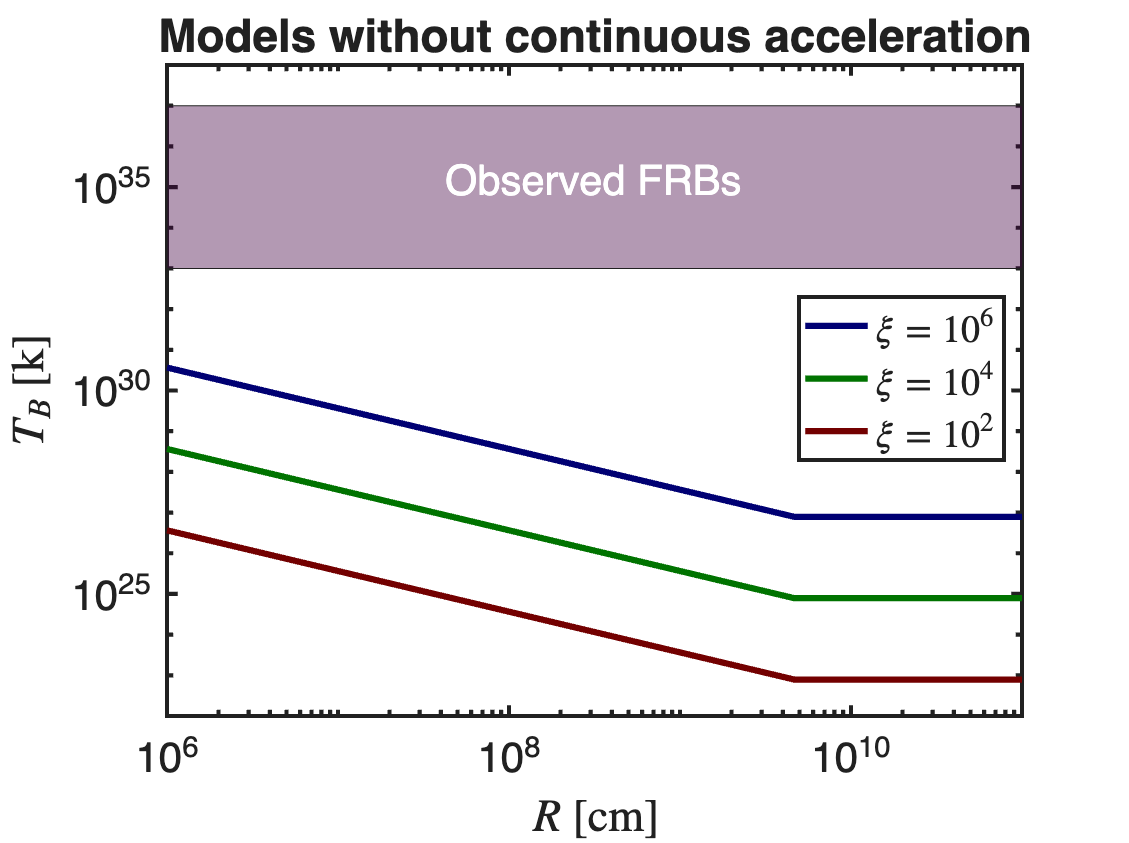}
\includegraphics[width=0.5\textwidth]{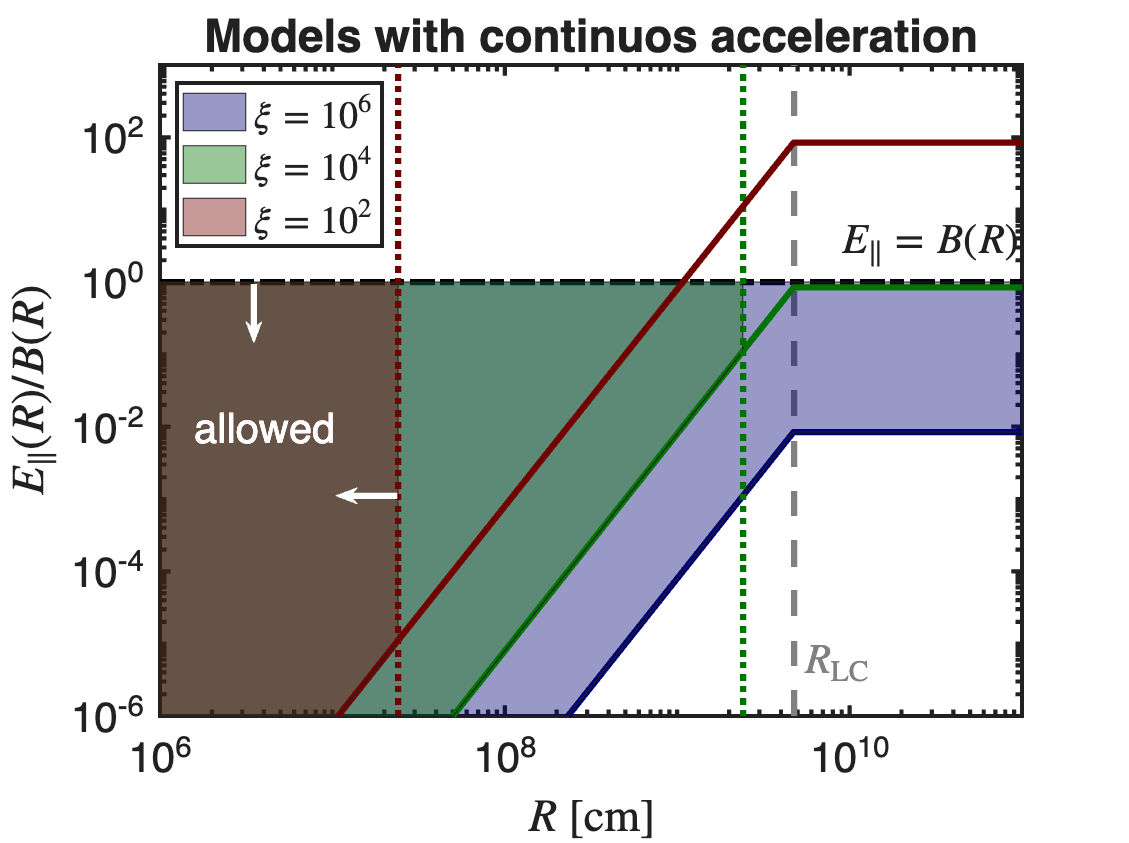}
\caption{Top: Observationally inferred brightness temperature as a function of the radius of the plasma beam for models involving no \emph{in situ} acceleration  (Eq. \ref{Tb-max}). 
Magnetospheric (\S \ref{sec:magnetospheric}) and intermediate (\S \ref{sec:intermediate}) models without continuous acceleration, fall short by orders of magnitude as compared with the observed values ($T_{B}\sim 10^{33}-10^{37}$K, shown by a shaded region). Bottom: Radius constraints based on: (i) the central engine frame electric field strength required in order to sustain the acceleration in the same models ($E_{\parallel}<B(R)$, solid lines) and (ii) charge availability (\S \ref{sec:charge}, dotted lines). Only models with sufficiently small radii (minimum of Eqns. \ref{eq:Rcharge}, \ref{eq:Rmax}, depicted by shaded regions) can potentially support the necessary acceleration.
Parameters assumed for these estimates are: $P=1\mbox{ s}, \gamma_j=300, t_{\rm FRB}=1\mbox{ ms},\nu=10^9\mbox{ Hz},L_{\rm FRB}=10^{42}\mbox{erg s}^{-1},\varepsilon_{_{\rm rad}}=10^{-3}$ and three different values of $\xi=10^2,10^4,10^6$.
} 
\par\vspace{1em}
\label{fig:TBkineticluminosity}
\end{figure}

\subsection{Regime of applicability}
\label{sec:applicable}
All neutron-star engine timescales are $\gg\nu_{\rm FRB}^{-1}$ (top of \S \ref{sec:GeneralBottleneck}). As a result, the GHz mode associated with FRB radiation requires generation of this mode during the process generating the bursts. A natural way to do this is by having an oscillatory current with a large Fourier amplitude at the frequency of the emerging daughter mode in the local emission frame, ${\bf J}_{\omega'}'$. If this current is also the energetic source of the radiation then $E_{\rm FRB}'=|\int {\bf J}_{\omega'}' \cdot {\bf E}_{\omega'}'dV'dt'|$ (primes denote quantities in the local plasma rest frame). Thus, the integrated energy that passes through particles equals the radiated energy, even if their instantaneous kinetic-energy content is much smaller because their energy is rapidly replenished (as indeed required for FRBs and detailed in \S \ref{sec:beam} - a similar situation arises also in a simple LC circuit). This includes the usual antenna picture: the energy can reside predominantly in macroscopic electric or magnetic fields, rather than in particle kinetic energy, but the oscillating current transfers this energy into the outgoing mode, and the corresponding integrated work is of order the radiated energy. Only if the current acts predominantly as a reactive mediator - effectively a catalyst transferring energy directly between electromagnetic degrees of freedom with little net particle energization - can this conclusion be avoided. This however, is subject to the conservative-Manley-Rowe \citep{1956PIRE...44..904M} conversion constraints discussed below and in more detail in \S \ref{sec:conserve}.

Such nearly conservative wave-wave conversion can occur when reactive plasma currents transfer energy from a pre-existing electromagnetic pump into the GHz daughter while acquiring little secular kinetic energy themselves.
Direct conversion from a low-frequency pump to $\omega_{\rm FRB}/\omega_0\gg 1$ is an extremely high-order process in a weakly nonlinear medium. Sequential resonant interactions or a strongly nonlinear many-mode cascade can avoid this high-order suppression and cannot be excluded model-independently. That said, such a mechanism must transfer power over many decades in frequency while satisfying the relevant resonance conditions and ultimately selecting an escaping GHz mode. The latter requirement is non-trivial because strong nonlinear cascades naturally produce broad spectra, whereas many FRBs exhibit narrow instantaneous spectral structure. We discuss the corresponding phase-matching conditions in \S \ref{sec:conserve}.

A related way to bypass the kinetic-energy bottleneck is to exploit a relativistically moving structure, so that the frequency increase is kinematic rather than arising from high-order frequency conversion. Relativistic IC scattering provides one familiar example. That being said, unless the mirror is highly magnetically dominated, it will itself posses a kinetic energy at least as large as that of the FRB.
A concrete example of this idea was recently presented by \cite{BD2026}. In their model an electromagnetic explosion pulse accelerates the magnetospheric plasma to $\Gamma\gg1$. Ambient kHz waves are Doppler shifted to MHz frequencies in the local fluid frame, where a conservative three-wave process $F\rightarrow F+A$ amplifies new F-waves. The latter are observed at $\nu_{\rm FRB}\sim\Gamma^2\nu_0\sim{\rm GHz}$. The large frequency shift therefore happens without an extremely high-order nonlinear interaction. The additional photon energy is extracted primarily from the electromagnetic explosion pulse rather than from particle kinetic energy, providing a genuine exception to the kinetic-energy bottleneck described above. The plasma is nevertheless required to supply the currents that mediate the interaction, and the model explicitly requires the density to remain above the charge-starvation threshold. Thus, while conservative relativistic wave conversion may evade the kinetic-energy requirement, it does not evade the independent charge-availability constraint discussed in \S \ref{sec:charge}.


\subsection{Charge availability}
\label{sec:charge}
As described in \S \ref{sec:applicable}, while the requirements for this are non-trivial, the kinetic-energy argument can in principle be evaded by conservative wave conversion. A more basic constraint follows directly from the finite current that the available charges can supply. Let ${\bf J}'_{\omega'}$ denote the material-current component responsible for generating the first substantial electromagnetic component at $\omega_{\rm FRB}$. Coherent generation of an EM wave over a radial depth $\ell'$ requires $\delta B_{\rm FRB}'\lesssim(4\pi/c)J'_{\omega'}\ell'$. Charge availability provides an upper limit on the current, $J'_{\omega'}\le en'c$. Plugging back into $\delta B'_{\rm FRB}$,
\begin{equation}
\label{eq:chargeavailable}
   n'\ell'\gtrsim{\delta B'_{\rm FRB}\over4\pi e} \to n=\gamma n'\geq \frac{L_{\rm FRB}^{1/2}}{4\pi e c^{1/2} R\ell'}. 
\end{equation}
where we have used $\delta B_{\rm FRB}^2R^2c\gtrsim L_{\rm FRB}$ and up to factors of order unity $\delta B_{\rm FRB}\approx \gamma \delta B_{\rm FRB}'$.
This constraint is independent of whether the current performs net work and therefore applies also to purely reactive frequency conversion. Taking the maximally available causally connected region $\ell'\lesssim R/(2\gamma_{\rm j})$ (see \S \ref{sec:beam}), and comparing with $\xi n_{\rm GJ}$, we get
\begin{equation}
\label{eq:Rcharge}
    R\lesssim 1.1\times 10^{10}\xi_5 B_{0,15} \Omega_1 L_{\rm FRB,42}^{-1/2}\gamma_{\rm j,3}^{-1}\mbox{ cm}.
\end{equation}
As will be shown in \S \ref{sec:magnetospheric}, this limit is comparable and consistent with limits obtained by applying the kinetic bottleneck constraint to magnetospheric models. Any shorter coherence length, incomplete participation of the available charges, or suppression of the relevant current by magnetization can only strengthen this bound.

\section{Application to FRB models}
Many FRB models invoking particle jets, do not include the required acceleration and therefore cannot account for the observed $t_{\rm FRB},T_B$ and $L_{\rm FRB}$ (i.e. $L_{\rm rad}$ in Eq. \ref{eq:Lmax}). In this section we consider in detail the implications of these constraints on different proposed models. The expected inferred brightness temperature for different model classes and the allowed radii in which continuous acceleration can be supported are shown in Fig. \ref{fig:TBkineticluminosity}. The magnetic field is taken as dipolar $B\propto R^{-3}$ up to the light-cylinder, and as $R^{-1}$ above. The particle density is $\xi n_{\rm GJ}(R)$ within the magnetosphere and falls off as $n\propto R^{-2}$ above the light-cylinder (see e.g. \citealt{Kirk2009}). 

Our results outlined in the following subsections are summarized in Table~\ref{tab:summary}. First, inner-magnetospheric models (\S \ref{sec:magnetospheric}) fail on energetics unless in-situ acceleration is invoked, at which point they survive only at sufficiently small radii and require a specific microphysical bunching driver.
Second, forced-reconnection models (\S \ref{sec:intermediate}) pass the acceleration test automatically but face a tight energy budget / low efficiency crisis. Third, external shock maser models (\S \ref{sec:faraway}) have ample energy at large radii, but the same parameters that make those radii energetically accessible also make the upstream wind optically thick to induced Compton scattering, preventing the emitted radio waves from escaping, without additional modifications that lead to further tension with observations (either in terms of spectral width or required efficiency). A hybrid model, synchrotron maser from monster shocks, magnetospheric in origin but relying on maser emission from a shock, is addressed separately in the table and discussed in \S \ref{sec:monster}.

\begin{table*}
\centering
\footnotesize
\caption{Summary of binding constraints on the three main FRB emission-model classes. Each class faces a different fundamental limitation; none of the classes survive without specific microphysical assumptions. Values are evaluated at fiducial FRB parameters: $L_{\rm FRB} \sim 10^{42}\mbox{erg s}^{-1}$, $t_{\rm FRB} \sim 1$\,ms, $\nu_{\rm FRB} \sim 1$\,GHz.}
\label{tab:summary}
\begin{tabular}{c c c c}
\hline\hline
\textbf{Model class} & \textbf{Binding constraint} & \textbf{Resulting limit} & \textbf{Required to survive} \\
\hline
Magnetospheric w/o in-situ acc. & $T_{\rm B}$ ; $L_{\rm rad}$ ; $t_{\rm c,obs}$ vs.\ $T$ & Falls short of FRB $T_{\rm B}$ and $L_{\rm FRB}$ & Continuous in-situ re-acceleration \\
 & (Eqns.~\ref{Tb-max}, \ref{eq:Lmax}, \ref{eq:tcmax}) & by $\gtrsim 7$ and $\gtrsim 5$ orders of magnitude & \\
\hline
Magnetospheric w. in-situ acc. & $E_{\parallel}\!\lesssim \!B(R)$ ; $N_{\rm max}$ ;$J'\!\leq\! n'ec$ & $R\lesssim 10^{10}$\,cm (any mechanism) & Transversely coherent driver \\
 & (Eqns. \ref{eq:Rmax}, \ref{eq:Nscmax}, \ref{eq:Nexmax}, \ref{eq:chargeavailable})  &  & of size $\gtrsim \sqrt{R\lambda}$ \\
\hline
Magnetospheric monster shock & Single-pass particle energy budget  & Extreme density / high $\nu_{\rm peak}$  & FMS pulse must clear magnetosphere ahead of precursor; \\
 & radiative-shock efficiency uncertain & if particle-limited  & (tension with observed $L_{\rm FRB}, \nu_{\rm FRB}$) \\
\hline
Light-cylinder reconnection & $\epsilon_{\rm iso},\varepsilon_{\rm rec,tot}$ & Falls short of FRB energies & Contradicting and extreme requirements  \\
 & (Eq.~\ref{eq:Eisorec}, \ref{eq:recefficiency}) & and extremely inefficient & to produce energetic FRBs with plausible efficiency \\
\hline
External shock maser & $\tau_{\rm ICS}$ ; $\varepsilon_{\rm tot}$ & Severely Compton-thick & High $\nu$ tail above peak (tension w. narrow-band FRBs); \\
 & (Eqns.~\ref{tau-IC-frb7},\ref{eq:epstot}) & Low efficiency & or $\sigma_{\rm w} \gtrsim 30$ (suppresses efficiency further) \\
\hline
\end{tabular}
\end{table*}

\subsection{Magnetospheric models}
\label{sec:magnetospheric}
As outlined in \S \ref{sec:beam}, in-situ acceleration is necessary in order for magnetospheric models to achieve the high brightness temperature and luminosities observed in FRBs. This requires a sufficiently strong electric field, in which particles can be accelerated. Using Eq. \ref{eq:tcmax}, we can estimate the (central engine frame) electric field strength along the particles' trajectories which is presumed to be along the background magnetic field:
\begin{equation}
\label{eq:Eparallel}
    E_{\parallel}\sim \frac{\gamma_j m_e c^2}{e c t_{\rm c,CE}}\gtrsim 5.2 \times 10^2 \mbox{ esu }\Omega_1^{-1} B_{0,15}^{-1} \xi_5^{-1}L_{\rm FRB,42}.
\end{equation}
This is independent of radius.
The electric field inferred above brings the particles close to a radiation-reaction-limit, $eE_\parallel c\simeq P_{\rm rad}$. Thus the assumed particle distribution is generally not externally fixed: radiation reaction acts on a timescale comparable to the cooling time and can strongly reshape the energy and pitch-angle distribution. For loss rates increasing with particle energy, this tends to drive particles toward a narrow radiation-reaction equilibrium, and is in principle consistent with a narrow resulting spectrum. Nevertheless, emission mechanisms requiring a specific non-equilibrium phase-space distribution must demonstrate that this distribution can be continuously regenerated in the presence of such strong radiative back reaction. We therefore regard Eq. \ref{eq:Eparallel} as a necessary energetic condition, not a sufficient condition for a self-consistent emission model.
The background magnetic field, is $B(R)\gtrsim E_{\parallel}$. 
This is a conservative upper bound from energy conservation: since $E_{\parallel}$ is most likely sourced by the magnetic field, $E_{\parallel}^2<B(R)^2$ requires the parallel field's energy density not to exceed that of the magnetic field. At $E_{\parallel}\to B(R)$, force-free dynamics breaks down on timescales $\sim \omega_p^{-1}\ll t_{\rm FRB}$ and the field cannot be sustained. At smaller $E_{\parallel}$, sustained acceleration is in principle possible if the screening (via pair cascades) is balanced by the source (e.g., charge starvation, reconnection, twist).
Considering a dipolar field geometry, this places an upper limit on the emission radius of $B_0(R_0/R)^3>E_{\parallel}$ leading to 
\begin{equation}
\label{eq:Rmax}
    R<1.2\times 10^{10}B_{0,15}^{2/3}\Omega_1^{1/3}\xi_5^{1/3}L_{\rm FRB,42}^{-1/3}\mbox{ cm}.
\end{equation}
At such radii, the plasma in the magnetosphere can in principle account for the required electric field to accelerate the electrons. 
This is the case in the low-twist scenario of \cite{Wadiasingh2019}, where the magnetar field is sufficiently undisturbed to permit charge-starved gap formation that sustains strong field-aligned electric fields, accelerating particles and triggering pair cascades and generation of coherent radio emission.
Electric fields which sustain the acceleration and the coherent curvature emission are also required in the model explored by \cite{KumarBosnjak2020}, via charge starved Alfv\'en waves. Similarly, a large electric field is required in the model by \cite{Zhang2022}, whereby FRBs are created by IC scattering of kHz fast-magnetosonic waves (the existence of such kHz waves is well-motivated, see \citealt{Blaes1989,TD1995,Timokhin2008,2011MNRAS.410.1036V,QuZhang2024,Burnaz2025,Qu2026}), although the origin of the electric field is not specified in this case.

Other proposed magnetospheric models such as free-electron masers \citep{Lyutikov2021} do not include in-situ acceleration. Such models are subject to the energy supply constraints derived in \S \ref{sec:beam}, and are severely challenged to self-consistently reproduce FRB properties.

A potential concern regarding the framework presented here is that first-principles simulations of pulsar polar-cap discharges \citep{2010MNRAS.408.2092T,2020PhRvL.124x5101P} have demonstrated that pulsar radio emission may arise from non-stationary pair cascades and electromagnetic waves excited during plasma-screening events, rather than from steady beams or coherent particle bunches as assumed in classical pulsar emission models. We note, however, that our framework remains applicable to such mechanisms. First, our analysis explicitly shows that the pulsar coherent-emission mechanism cannot be straightforwardly extrapolated to FRB luminosities; the qualitative differences in brightness temperature and required particle number imply that even a fully-realistic theory of pulsar radio cannot automatically be adopted for FRBs. Second, and more fundamentally, in any such mechanism the radio output is ultimately powered by particle kinetic energy: the non-stationary electric field accelerates particles, which trigger pair cascades and excite plasma waves through screening dynamics. Energy continues to pass through the kinetic channel before being released as coherent radio, so the kinematic bounds derived above apply regardless of the specific coherence mechanism. 

An additional limitation on magnetospheric models is that the number of emitting particles per coherent bunch should not exceed $N_{\rm ex,max}$ as given by Eq. \ref{eq:Nexmax} (or the more restrictive $N_{\rm sc,max}$ as given by Eq. \ref{eq:Nscmax} for models that rely on self-coordination between particles to achieve coherence). Recalling that $P_{\rm em}>L_{\rm rad}/(2\pi n R^3 N_{\rm max})$ (Eq. \ref{eq:tcmax}), and combining with Eq. \ref{eq:Nexmax} we find that the minimum emission power per particle is required to be
\begin{eqnarray}
\label{eq:Pemmin}
  &  P_{\rm em}>\frac{3 L_{\rm rad}}{8\pi n^2 R^4 \lambda^2}\approx \\ & 2.7\times 10^{-19}L_{\rm FRB,42} \xi_5^{-2} R_9^2 B_{0,15}^{-2} \Omega_1^{-2} \nu_9^2\mbox{ erg s}^{-1}. \nonumber
\end{eqnarray}
The impact of the constraint given by Eq. \ref{eq:Pemmin} is model-dependent. For instance, in the IC scattering model, the emitted power per particle is orders of magnitude larger than given by Eq. \ref{eq:Pemmin} \citep{Zhang2022}, and this particular consideration is not informative, as opposed to, say, the constraint given by Eq. \ref{eq:Rmax}. As we show next, the situation is different in the case of the antenna mechanism of coherent curvature radiation.
For this mechanism, the emission frequency is $\nu\approx c\gamma_j^3/(2\pi R)$ where $R$ is the magnetic field lines' curvature radius, and we take it to be comparable to the emission radius. This leads to a required LF $\gamma_{j}\approx 590 \nu_9^{1/3}R_9^{1/3}$. The emitted power per particle for bunches moving in a curved trajectory is 
\begin{equation}
    P_{\rm cur}=\frac{2 e^2 \gamma_{j}^4 c}{3 R^2}\approx 5.7\times 10^{-16}\nu_9^{4/3}R_9^{-2/3}\mbox{ erg s}^{-1}.
\end{equation}
Combining with Eq. \ref{eq:Pemmin}, we find that coherent curvature radiation is consistent with the maximum number of particles per bunch as long as\footnote{A closely related constraint on coherent curvature radiation was derived by \cite{Cooper2021} who imposed the requirement that the bunch's self-induced perpendicular field not destabilize the coherence, yielding a maximum luminosity $L_{\rm coh}^{\rm max}\propto R^{-6}$. Their bound on the maximum coherent radius is consistent in spirit with our Eq. \ref{eq:Rmaxcurv}. The constraints differ in which physical effect sets the bunch-size limit (the self-induced $B_{\perp}$ for \cite{Cooper2021}; the energy budget and brightness temperature for us), but they yield bounds of similar magnitude and lead to the same qualitative conclusion: coherent curvature radiation at FRB luminosities requires emission radii relatively close to the stellar surface.}
\begin{equation}
\label{eq:Rmaxcurv}
    R<1.7\times 10^{10}\xi_5^{3/4} B_{0,15}^{3/4} \Omega_1^{3/4} L_{\rm FRB,42}^{-3/8} \nu_9^{-1/4}\mbox{ cm}.
\end{equation}
For curvature radiation, this constraint is comparable to Eq. \ref{eq:Rmax}, depending somewhat on the values of the model parameters and observables considered.

\subsubsection{Magnetospheric shock-maser mechanisms: monster shocks}
\label{sec:monster}
A recently developed scenario that merits separate comment is the `monster shock' model \citep{Chen+22,Beloborodov2023,VanthieghemLevinson25}, in which a fast-magnetosonic wave launched by a magnetospheric disturbance steepens into a strong radiative shock within the magnetosphere with coherent GHz emission produced as a precursor wave.  This model occupies an interesting hybrid situation in the framework described in this work. 
The shock-maser mechanism has been validated by multiple independent simulations beyond the force-free approximation, including local 1D PIC \citep{Chen+22,VanthieghemLevinson25}, global 2D PIC in a dipolar magnetosphere \citep{Bernardi+25}, and global 3D relativistic MHD \citep{Grehan2026}. These simulations confirm shock formation through nonlinear wave steepening and, at the kinetic level, the production of a coherent precursor wave through the synchrotron maser instability. The basic shock-formation physics is therefore on firm ground. Three questions, however, remain open from the perspective of FRB application: (i) whether the precursor wave can escape the surrounding magnetospheric pair plasma to reach the observer; (ii) whether the model can reproduce FRB-level luminosities; and (iii) the angular distribution of emission for non-equatorial shock geometries. The first and third questions are explicitly flagged as open in the simulation literature. Below we address the second, and in the process find that the same density requirements that would source FRB-level luminosities also exacerbate the escape problem.

The shock provides continuous acceleration in situ, and the constraints of \S \ref{sec:beam} are nominally satisfied.
However, the acceleration in the monster shock model relies on a continuous fresh supply of upstream material crossing the shock front. 
If the coherent emission from each particle is limited by the kinetic energy it carries into the shock, then sustaining the FRB luminosity over $t_{\rm FRB}$ requires at least the particle number given by Eq. ~\ref{eq:Nreq}. Under this assumption, Eq.~\ref{eq:Nmag} implies a density exceeding the Goldreich–Julian value by $\sim10^{10}$.
This raises two concerns. First, the escape problem. The required pair plasma can in principle be produced by the fast-magnetosonic wave itself through pair cascades, but this raises the local density by a vast factor in the very region the radio wave must traverse to escape, exacerbating the efficiency / induced-Compton opacity problem (as well as stimulated Raman scattering) analyzed in the context of the external shock case in \S \ref{sec:inducedCompton}. The escape problem for monster-shock FRBs therefore remains unresolved in this model, as the parameters required to make the emission viable are subject to conflicting demands: a higher particle density is needed to power the observed luminosity, whereas a lower particle density is required for the radiation to escape.
Second, the frequency problem: even if the required particle number can be supplied, the result is that the emission will peak at $\nu\gg 1$\,GHz. This is because, using Eq. \ref{eq:Nreq}, and taking the emission radius to be where the wave electric field is $|E|=B(R)/2$, the emission radius is $R\sim 6\times 10^7 B_{0,15}^{1/2}L_{\rm fms,45}^{-1/4}$\,cm (where $L_{\rm fms}\sim \varepsilon_{\rm rad}^{-1}L_{\rm FRB}$ is the luminosity of the fast magnetosonic wave). The required density is $n_{\rm req}\sim 10^{19}L_{\rm FRB,42}^{7/4}t_{\rm FRB,-3}\gamma_{j,5}^{-1}\varepsilon_{\rm rad,-3}^{-7/4}B_{0,15}^{-3/2}\mbox {cm}^{-3}$, where we have conservatively normalized the total radiative efficiency by $10^{-3}$, but we note that the true efficiency may be smaller, especially for a magnetized upstream (see \S \ref{sec:magshock}). Therefore, in the far upstream, the lab-frame plasma frequency is $\omega_{\rm p}\sim  1.7\times 10^{14} L_{\rm FRB,42}^{7/8}t_{\rm FRB,-3}^{1/2}\gamma_{j,5}^{-1/2}\varepsilon_{\rm rad,-3}^{-7/8}B_{0,15}^{-3/4}\mbox{rad s}^{-1}$. \cite{VanthieghemLevinson25} argue that in the immediate upstream of the monster shock, the plasma has been pre-accelerated by the wave to a large LF relative to the shock, $\gamma_{u}\sim 10^5$. The observed maser emission then peaks at $\omega_{\rm peak}\approx \sqrt{\sigma_u/\gamma_u} \omega_{\rm p}\sim 5\times 10^{12}\mbox{rad s}^{-1}$, where we have taken $\sigma_u\sim 94 B_{0,15}^{1/2}L_{\rm fms,45}^{-1/4}$ as in \cite{VanthieghemLevinson25}. This frequency is 5000 times larger than the GHz band in which FRBs are typically observed. These considerations suggest that, barring other difficulties, the monster shock mechanism may only be capable of producing very low-energy FRBs, but is unlikely to account for a typical FRB.
A recent calculation by \cite{VL2026}, suggests that strongly radiative monster shocks may instead convert a fraction $\sim 10^{-3}$ of the incident FMS-wave energy directly into the coherent precursor. In this case, particles crossing the shock can be continuously energized by the shock fields during their transit, and the single-particle energy budget assumed above need not apply. Their calculation nevertheless relies on a one-dimensional treatment and weak precursor-shock coupling; the nonlinear feedback and subsequent escape of the wave remain open questions. We therefore retain the above estimate as the constraint applicable when the radio output is limited by the kinetic energy carried into the shock.
After completion of this analysis, \cite{Beloborodov2026} posted a detailed self-consistent treatment of the monster shock and its precursor. Based on arguments distinct from those given here, that work reaches a strikingly similar conclusion: monster shocks inside the magnetosphere can produce FRBs only at the low-energy end of the population, comparable to the SGR 1935+2154 event. Moreover, increasing the energy of the launching fast-magnetosonic wave strongly reduces the maser efficiency, ruling out the model for typical cosmological FRBs regardless of the input energy.

 \subsubsection{Wave escape from the magnetosphere}
 \label{sec:escape}
We have identified in this section conditions in which the constraints related to the kinetic energy bottleneck and the charge starvation can be satisfied. Nonetheless, this does not yet guarantee that such models are viable ways of explaining the FRB phenomenology. An important unresolved aspect is whether the FRB waves can escape the magnetosphere. We provide here a brief overview of this topic.

\cite{Beloborodov2021}, argued that escape is strongly inhibited by enhanced scattering and dissipation once the background magnetic field becomes comparable to the magnetic field of the wave itself.
\cite{Beloborodov2023,Beloborodov2024,VanthieghemLevinson25} suggest that high-amplitude GHz waves steepen as they propagate to larger radii in the magnetosphere because the refractive index depends on the magnetization parameter. The waves eventually develop into shocks at the radius where the wave magnetic field amplitude reaches half the strength of the magnetar's dipole magnetic field. These shocks can lead to efficient dissipation of the wave energy.

Other studies (e.g., \citealt{Qu-etal2022,Lyutikov2024,QuKumar2026}) have reached a different conclusion. They suggest that FRB waves can escape along open magnetic field lines and may even clear a path for their own propagation by accelerating and sweeping aside the surrounding plasma. This process could be driven by the strong ponderomotive force exerted by the leading edge of the FRB pulse. Induced Compton and stimulated Raman scattering could further facilitate plasma acceleration. Indeed, if an FRB pulse were to dissipate at radius $r$ in the magnetosphere and transfer its momentum to the plasma, it could accelerate the particles to a Lorentz factor of\footnote{This estimate assumes that the magnetic field remains anchored to the neutron star and is not advected with the plasma flow.} $\sim 10^6 \xi_6^{-1} L_{\rm FRB,42}t_{\rm FRB,-3}$. Thus, the leading edge of an FRB pulse could plausibly clear a path through the magnetosphere, allowing the remainder of the pulse to propagate with little impediment. A weak precursor or the leading edge of a bright pulse may similarly accelerate the ambient plasma to highly relativistic speeds along the magnetic field lines, thereby clearing a channel through which the remainder of the burst can propagate freely. Moreover, the dipolar magnetic-field geometry commonly assumed in escape calculations that find high-amplitude GHz waves unable to escape the outer magnetosphere may not apply during an FRB, when the magnetosphere is expected to be strongly disturbed. Alfv\'en waves launched during the event that produces an FRB can open up the magnetic field lines, producing a configuration closer to a split monopole (e.g., \citealt{Most2024,Burnaz2025}). As the field geometry transitions from dipolar to split-monopolar, the background magnetic field may remain strong enough to keep the EM wave amplitude below it out to the light cylinder. This could suppress the wave-scattering and dissipation mechanisms proposed to operate within the magnetosphere.

Thus, while current observations favor compact near-engine emission, there are theoretical questions and issues regarding the wave escape from the magnetosphere that continue to be investigated and debated. 

\subsection{Intermediate, light-cylinder, model - Forced reconnection}
\label{sec:intermediate}
In forced reconnection models \citep{Philippov+2019,Lyubarsky2020}, the energy is carried away from the NS in the form of Poynting flux, which is subsequently converted via reconnection into kinetic energy of the particles before a fraction is released as radio waves. A crucial distinction from beam-driven models is that the reconnection layer itself serves as a continuous particle acceleration site, so the kinetic energy need not be transported to a separate emission region. The constraints derived in \S\ref{sec:beam} therefore do not apply directly to this class of models. 

Crab nanoshots are the only known transients with comparable brightness temperatures  to FRBs, $T_{\rm B,ns}=10^{37}-2\times 10^{41}\mbox{ K}$ \citep{Hankins2003,Hankins2007,Jessner2010}, making them a particularly instructive comparison. A version of the forced reconnection model, operating outside the light cylinder, has been proposed to power these nanoshots \citep{Philippov+2019}.
We note that this interpretation of the giant-pulse emission has been questioned by alternative proposals \citep{2024ApJ...974L..21W} and by the substantial X-ray energetics associated with giant pulses \citep{2021Sci...372..187E}, which may point toward an inner-magnetospheric origin instead. Our analysis below is agnostic to the correct picture for Crab nanoshots; we consider only the consequences of extrapolating the mechanism to FRB luminosities, which we show face serious difficulties independent of the nanoshot question.
Taking the relevant Crab observables - $B_0\approx 10^{12}$\,G, $\Omega\approx 190\mbox{ rad s}^{-1}$, $\nu_9\approx 3-10$, $t\approx 0.4-4\times 10^{-9}$\,s, as well as $R\approx R_{\rm LC}=c/\Omega\sim 1.6\times 10^8\mbox{ cm}$, $\gamma_j\approx 10^4$, $\xi\approx 10^2$ as required by the reconnection model - we can evaluate the beam-driven bounds derived above and ask whether they are satisfied.
The answer is negative. Applying these parameters to Eq.~\ref{Tb-max} gives $T_{\rm B}\approx \varepsilon_{_{\rm rad}} 10^{35-38.5}{\rm K}$, inconsistent with the observed values at the upper end of the nanoshot brightness temperature range. The luminosity bound from Eq.~\ref{eq:Lmax} yields $L_{\rm rad}\approx \varepsilon_{_{\rm rad}} 8\times 10^{33}\mbox{erg s}^{-1}$ falling well below the observed peak luminosities of $L_{\rm rad,ns}\approx 10^{38-39}\mbox{erg s}^{-1}$. The cooling time estimate is the most vivid illustration: Eq.~\ref{eq:tcmax} gives $t_{\rm c,ns}\lesssim 2\times 10^{-15}\,$s, up to six orders of magnitude shorter than the nanoshot duration.

These three estimates establish that one-shot beam-driven emission is untenable even for the Crab nanoshots. The reconnection model of \citet{Philippov+2019} evades all three bounds for the same reason: the reconnection layer continuously replenishes the particle energy at the emission site. 
However, note that the acceleration must operate on a very short timescale. It should refresh the particles' energy on an observed timescale of $t_{\rm c,ns}\sim 2\times 10^{-15}$\,s, orders of magnitude shorter than the dynamical timescale.
The Crab nanoshots thus serve as direct demonstration of the conclusion that continuous in-situ acceleration (provided by reconnection or otherwise) is a necessary ingredient for producing the most extreme coherent radio transients.

Moving from the Crab nanoshots to FRBs, we can once more place a lower limit on the required electric field to continuously accelerate the particles. Two additional considerations modify the limit provided by Eq. \ref{eq:Rmax}. First, in reconnection models the electric field is $E_{\rm rec}\sim \beta_{\rm in} B_{\rm rec}$ with $\beta_{\rm in}\sim 0.1$ being the normalized velocity of particles flowing into the reconnection layer. Second, the wind magnetic field at the light-cylinder, $B_{\rm w}(R_{\rm LC})$, is compressed by the fast magnetosonic (FMS) pulse by a factor of $\sim B_{\rm fms}/B_{\rm w}(R_{\rm LC})$, such that the reconnecting field is $B_{\rm rec}\sim B_{\rm fms}$
where
\begin{equation}
    B_{\rm fms}=\sqrt{\frac{L_{\rm fms,47}}{cR_{\rm LC}^2}}\approx 6\times 10^8L_{\rm fms,47}^{1/2}\Omega_1\mbox{ G}
\end{equation}
is the magnetic field associated with the FMS pulse. 
The width of the compressed wind at the light cylinder is $\sim R_{\rm LC} B_{\rm w}(R_{\rm LC})/B_{\rm fms}$. If, at the point of reconnection, the wind is moving outwards at a LF $\Gamma_{\rm w}$, then the pulse can only overtake a region narrower than this by a factor $2\Gamma_{\rm w}^2$ before the wind expands significantly \citep{BK2023}, so that the effective width is $W_{\rm rec}\sim R_{\rm LC} B_{\rm w}(R_{\rm LC})/(2\Gamma_{\rm w}^2B_{\rm fms})$. The radiated energy in this model, is a fraction $\varepsilon_{\rm rad}$ of the magnetic energy in the reconnecting layer \footnote{For $\Gamma_w \to 1$ (a mildly relativistic wind at the reconnection site near the light cylinder, as adopted in \citealt{Lyubarsky2020}), Eq. \ref{eq:Eisorec} recovers his estimate $\epsilon \sim B_{\rm pulse} B_{\rm w} R_{\rm LC}^3$. For larger $\Gamma_w(R_{\rm LC})$, the available reconnection energy is further suppressed by $2\Gamma_w^2$, making the energy budget even tighter.}:
\begin{eqnarray}
\label{eq:Eisorec}
   & \epsilon_{\rm iso}\sim \varepsilon_{\rm rad} 4\pi R_{\rm LC}^2 W_{\rm rec} \frac{B_{\rm fms}^2}{4\pi}\sim \varepsilon_{\rm rad} B_0 R_0^3 B_{\rm fms}/(2\Gamma_{\rm w}^2)\nonumber \\
   &\sim 3\times 10^{39} \varepsilon_{\rm rad,-2}B_{0,15} L_{\rm fms,47}^{1/2} \Omega_1\Gamma_{\rm w}^{-2}\mbox{ erg.} \nonumber \\
   & \sim 2\times 10^{39} \varepsilon_{\rm rad,-2} L_{\rm fms,47}^{1/2}\Gamma_{\rm w}^{-2}\bigg(\frac{30 \mbox{ days}}{t_{\rm sd}}\bigg)^{1/2}\mbox{ erg.}
\end{eqnarray}
where $t_{\rm sd}$ is the spindown time.
This equation shows that it is not trivial for this model to account for the isotropic equivalent energies of FRBs, which can reach $\sim 10^{42}\mbox{ erg}$ (e.g. \citealt{Shah2026}), even for an extremely short spindown time (which is clearly incompatible with repeaters detected observed for over a decade) and if the wind is only mildly relativistic at the light-cylinder $\Gamma_{\rm w}\approx 1$.

The observed burst duration in this model reflects that of the magnetic pulse launched from the NS surface. Therefore, we have
$L_{\rm FRB}\sim \epsilon_{\rm iso}/t_{\rm FRB}$. Plugging this into Eq. \ref{eq:Eparallel} and comparing with $E_{\rm rec}$, the electric field condition becomes:
\begin{equation}
\label{eq:Ereqrec}
    2.6\times 10^{-5}<\varepsilon_{\rm rad,-2}^{-1} t_{\rm FRB,-3} \Omega_1 \Gamma_{\rm w}^2\beta_{\rm rec,-1}\xi_5
\end{equation}
which can be easily satisfied for typical model parameters. 

While $\varepsilon_{\rm rad}$ gives the radiative efficiency, Eq. \ref{eq:Eisorec} shows that the total efficiency associated with this model is much lower due to the relatively small amount of usable magnetic energy as compared with the energy of the FMS pulse,
\begin{eqnarray}
\label{eq:recefficiency}
   & \varepsilon_{\rm rec,tot}\!=\!\frac{\epsilon_{\rm iso}}{N_{\rm fms}L_{\rm fms}t_{\rm FRB}} \\
   &\approx \! 3\times 10^{-6} \varepsilon_{\rm rad,-2}B_{0,15} L_{\rm fms,47}^{-1/2} \Omega_1\Gamma_{\rm w}^{-2}t_{\rm FRB,-3}^{-1}N_{\rm fms,1}^{-1} \nonumber \\
   & \approx 2\times 10^{-6} \varepsilon_{\rm rad,-2} L_{\rm fms,47}^{-1/2} \Gamma_{\rm w}^{-2}t_{\rm FRB,-3}^{-1} \bigg(\frac{30 \mbox{ days}}{t_{\rm sd}}\bigg)^{1/2}N_{\rm fms,1}^{-1}\nonumber \\
   & \approx 4\times 10^{-6} \varepsilon_{\rm rad,-2}^2 \Gamma_{\rm w}^{-4}t_{\rm FRB,-3}^{-1} \bigg(\frac{30 \mbox{ days}}{t_{\rm sd}}\bigg) \bigg(\frac{10^{39} \mbox{ erg}}{\epsilon_{\rm iso}}\bigg) N_{\rm fms,1}^{-1}\nonumber
\end{eqnarray}
where $N_{\rm fms}$ is the number of pulses released by the NS in a given emission episode. This factor accounts for the fact that only the first pulse can lead to forced reconnection, as it significantly disturbs the magnetosphere, and the latter ejected pulses, do not interact with an unperturbed striped wind \citep{BK2023}. 
Eq. \ref{eq:recefficiency} represents a tiny efficiency, even for an extremely young magnetar and when allowing for an only mildly relativistic wind at the light-cylinder, $\Gamma_{\rm w}\to 1$. Increasing $L_{\rm fms}$ raises $\epsilon_{\rm iso}$ but decreases $\varepsilon_{\rm rec,tot}$, at the same rate, so no choice of $L_{\rm fms}$, simultaneously raises the available energy and the conversion efficiency. 

\begin{figure}[ht]
\centering
\includegraphics[width=0.5\textwidth]{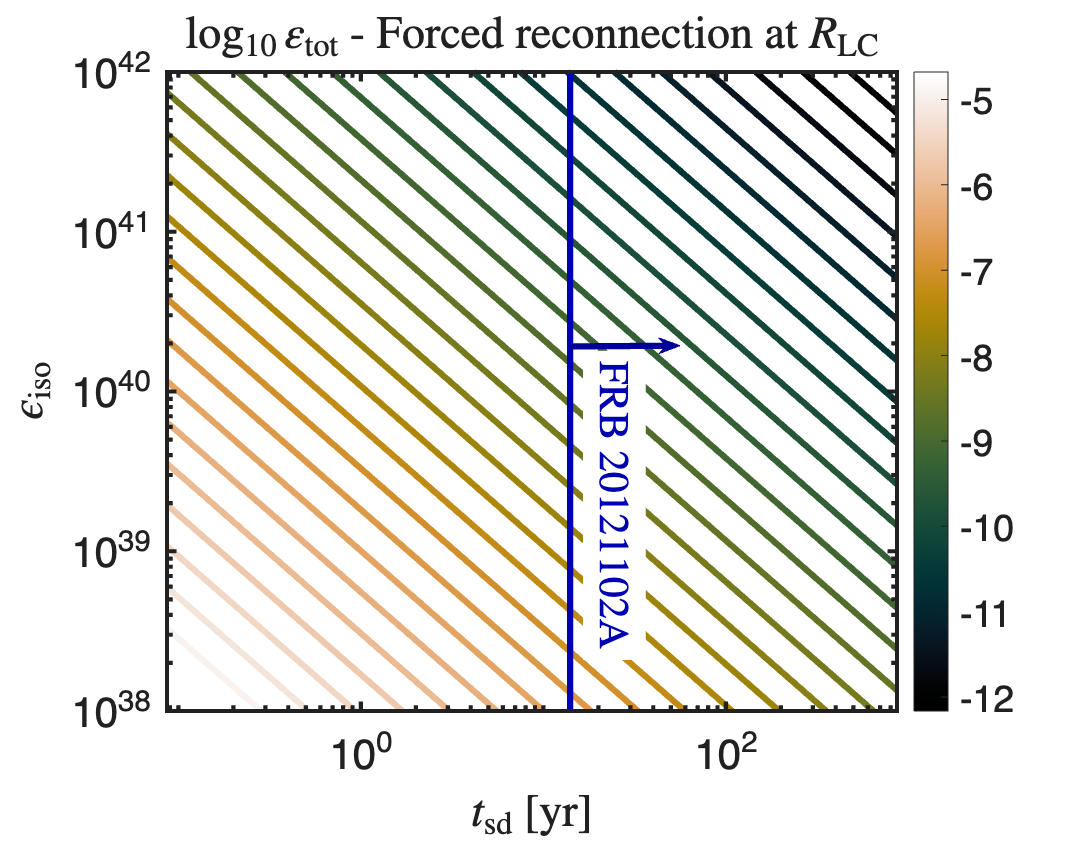}
\caption{Total efficiency in the forced reconnection model ($\varepsilon_{\rm tot}$) as a function of the isotropic equivalent FRB energy, $\epsilon_{\rm iso}$ and the assumed spindown age of the NS source, $t_{\rm sd}$. For FRB 20121102A, the true age is $>14$\,yr (vertical line) and this leads to tiny efficiencies in this model, especially for higher energy bursts. Results are plotted for a radiative efficiency (fraction of energy in reconnection zone, radiated in radio) of $\varepsilon_{\rm rad}=10^{-2}$ as well as for $\Gamma_{\rm w}=1, N_{\rm fms}=10,t_{\rm FRB}=1\mbox{ ms}$, see \S \ref{sec:intermediate} for details.
} 
\par\vspace{1em}
\label{fig:reconnect}
\end{figure}

To summarize, forced reconnection models naturally provide in-situ acceleration, and the corresponding electric-field condition is straightforwardly satisfied. The principal challenge for the FRB application is energetic: producing the observed isotropic-equivalent energies with reasonable efficiencies requires combinations of high $\varepsilon_{\rm rad}$, low $\Gamma_{\rm w}$, and very short $t_{\rm sd}$, beyond their plausibility limits. The efficiency of this model as a function of $\epsilon_{\rm iso}$ and $t_{\rm sd}$ is presented in Fig. \ref{fig:reconnect}.
In particular, this is very difficult to reconcile with active repeaters. 
For FRB 20121102A's age of $>14$\,yr, $t_{\rm sd}$ would need to be much shorter than 14 yr to power flares at the present epoch. This contradicts the system remaining continuously active at comparable levels since detection.
Other, independent, difficulties of the forced reconnection model of FRBs include the expected spectrum, which is wide as opposed to what is observed in many bursts \citep{Zhou2022,Zhang+2023,Sand2025}.

\subsection{Far away model - synchrotron maser}
\label{sec:faraway}
Synchrotron maser emission from relativistic shocks located at AU-scale distances from the central engine has been widely proposed as a mechanism for powering FRBs \citep{Lyubarsky14,Beloborodov17,Beloborodov19,Metzger+19,Margalit+20}. In this picture, a highly energetic baryonic outflow launched by the neutron star drives a relativistic shock into the surrounding medium, where coherent radio emission is produced.

As discussed in \S\ref{sec:intro}, several recent observations have been interpreted as favoring emission sites much closer to the central engine. However, these arguments are not yet conclusive, and far-away blast-wave models remain among the leading FRB scenarios. Moreover, the constraints developed in this work are independent of the emission location and can therefore be applied directly to such models. In this section we examine the implications of the kinetic-energy bottleneck for relativistic-shock FRB scenarios.

The injected luminosities invoked in these models are typically very large, $\sim10^{46}-10^{49},\mathrm{erg,s^{-1}}$ \citep{Margalit+20}. Combined with the large emission radii, $R\sim10^{12}-10^{13},\mathrm{cm}$, these parameters allow the beaming constraints derived in \S\ref{sec:beam} to be satisfied without requiring continuous particle acceleration. However, the same large energy fluxes and upstream densities that enable this also imply substantial induced-Compton opacity for coherent GHz radiation propagating ahead of the shock, as discussed next.

\subsubsection{induced Compton optical depth}
\label{sec:inducedCompton}

 If the FRB radio pulse is generated at $R$ via the shock-maser process, then the observed duration of the burst is:
\begin{equation}
    t_{\rm FRB} \approx {R\over 2c \Gamma_{\rm sf}^2}.
    \label{t-frb-shock-maser}
\end{equation}
where $\Gamma_{\rm sf}$ is the LF of the shock front. Following the setup proposed in \cite{Metzger+19} we focus on dissipation in the forward shocked region. We consider five different LFs involved in the shock interaction. In addition to $\Gamma_{\rm sf}$, these are: $\Gamma_w$ (the LF of the upstream wind), $\Gamma_{\rm sp}$ (the LF of the shocked plasma), $\gamma_u$ (the LF of the unshocked wind relative to the shock front), and $\gamma_d$ (the LF of the shocked plasma relative to the shock front). As shown below (Eq. \ref{gamu-gamd}), only two of these five LFs are apriori unknown. They depend on the LF and luminosity of the relativistic outflow driving the shock.

For a highly relativistic shock, the LF of the shocked plasma associated with its bulk motion relative to the shock front ($\gamma_d$) corresponds to that of the appropriate wave that transports signals downstream to within a factor of order unity. For an upstream wind magnetization $\sigma_{\rm w} \gg 1$, the relevant MHD wave propagates with the Alfv\'en speed and has a LF of $\sqrt{1 + \sigma_{\rm w}}$. In contrast, for $\sigma_{\rm w} \ll 1$, the signal propagates with the sound speed, which, for a highly relativistic shock, is $\sim c/\sqrt{3}$.
An equation for ($\gamma_d$) valid in both the ($\sigma_{\rm w} \gg 1$) and ($\sigma_{\rm w} \ll 1$) limits is\footnote{$\gamma_d\approx (\sigma_{\rm w} + 9/8)^{1/2}$ for $\sigma_{\rm w} \gg 1$ and $\gamma_d\approx \sqrt{9(1+\sigma_{\rm w})/8}$ for $\sigma_{\rm w} \ll 1$ (e.g. \citealt{narayan2011}). In the limit $\sigma_{\rm w}\rightarrow0$, $\gamma_d=\sqrt{9/8}$ or $\beta_d=1/3$. For a detailed description of MHD shocks, see \cite{Kennel1984}.}, $ \gamma_d \approx (\sigma_{\rm w} + 9/8)^{1/2}$. For ease of readability, we use below the slightly cruder approximation $\gamma_d \approx (\sigma_{\rm w} + 1)^{1/2}$ which simplifies the algebra. $\gamma_d$ and $\gamma_u$ are related to the other LFs via Lorentz transformations
\begin{equation}
    \gamma_d \approx {\Gamma_{\rm sf}\over 2\Gamma_{\rm sp}} \approx (\sigma_{\rm w} + 1)^{1/2}, \quad \gamma_u \approx {\Gamma_{\rm sf}\over 2\Gamma_w}.
    \label{gamu-gamd}
\end{equation}

The kinetic energy flux carried by electrons, as seen in the shock front comoving frame\footnote{\label{LT-flux} One factor of $\gamma_u$ in the expression for $F_{\rm e,sf}$ represents the relative LF of the upstream plasma with respect to the shock front. This corresponds to Lorentz contraction of length, meaning the density in the shock front frame is larger than in the CSM rest frame by this factor. Furthermore, the energy of each particle in the shock front comoving frame is $\sim \gamma_u m_{\rm w} c^2 $, which accounts for the other factor in the expression for $F_{\rm e,sf}$.}, and the total energy flux that includes the contributions from the magnetic field are:
\begin{equation}
    F_{\rm e,sf} = m_e c^3 n_{\rm w} \gamma_u^2 \quad{\rm and}\quad F_{\rm tot,sf} = m_{\rm w} c^3 n_{\rm w} (1+\sigma_{\rm w}) \gamma_{u}^2.
    \label{shock-energy-flux0}
\end{equation}
where $m_e$ is electron mass, $m_{\rm w}$ is the particle mass that dominates the CSM density and $n_{\rm w}$ is the particle density in the upstream wind frame. Here and in what follows $Q_{x,y}$ denotes the (central engine) quantity $Q_x$ as measured in frame $y$ (e.g. shock front frame in Eq. \ref{shock-energy-flux0}).

According to numerical simulations, a fraction $\varepsilon_{\rm ma}$ of the electron kinetic energy flux in the shocked plasma frame is converted to radio photons via the synchrotron maser process\footnote{$\varepsilon_{\rm ma}$ is equivalent to the parameter $g_{\xi}$ defined in  \cite{Plotnikov&Sironi19}, defined relative to the electron kinetic energy flux in the shocked plasma frame. This differs from their $f_{\xi}$ defined relative to the total incoming energy flux, by a factor of $(1+\sigma_{\rm w})$; in particular, $g_{\xi}$ remains approximately constant at high $\sigma_{\rm w}$, whereas $f_{\xi}\propto \sigma_{\rm w}^{-1}$.}. Thus, the maser photon energy flux in the rest frame of the shocked plasma  is $F_{\rm ma,sp} \approx \varepsilon_{\rm ma} F_{\rm e,sp}$, with $F_{\rm e,sp}=(\Gamma_{\rm sp}/\Gamma_{\rm sf})^2F_{\rm e,sf}=F_{\rm e,sf}/[4(1+\sigma_{\rm w})]$. The energy flux in EM radiation in the central engine frame just outside the shock front is:
\begin{equation}
    F_{\rm ma} \approx F_{\rm ma,sp} (4\Gamma_{\rm sp}^2)
    \approx \varepsilon_{\rm ma}\, m_e n_{\rm w} c^3 \frac{\Gamma_{\rm sf}^4}{16(1+\sigma_{\rm w})^2\Gamma_w^2},
    \label{maser-flux-lab}
\end{equation}
The factor $4\Gamma_{\rm sp}^2$ in this equation is of similar origin to the factor $\gamma_u^2$ described in footnote (\ref{LT-flux}), except that this time we are dealing with radiation moving approximately radially outwards in the shocked plasma frame (for $\sigma_{\rm w}\gg1$) and hence the appropriate transformation is a Doppler as opposed to a Lorentz transformation. The particle density in the CSM is related to the central engine frame isotropic equivalent luminosity of the electrons $L_e$,
\begin{equation}
\begin{aligned}
    L_e = 4\pi R^2 n_{\rm w} m_e c^3 \Gamma_w^2 \quad \to \quad
     n_{\rm w} = {L_e\over 4\pi R^2 m_e c^3 \Gamma_w^2}.
 \label{n-wind}
\end{aligned}
\end{equation}
Substituting this back into equation (\ref{maser-flux-lab}), we obtain the bolometric isotropic equivalent luminosity of maser emission from a shock
\begin{equation}
    L_{\rm FRB} = 4\pi R^2 F_{\rm ma} \approx \varepsilon_{\rm ma}\,L_e \left[ \frac{\Gamma_{\rm sp}}{\Gamma_w}\right]^4.
    \label{L-frb-shock}
\end{equation}

The frequency of maser emission in the observer frame is:
\begin{equation}
    \omega_{\rm FRB} \simeq 3\Gamma_{\rm sp} \,\omega_p \, (1+\sigma_{\rm w})^{1/2}\approx \frac{3\,e\,\Gamma_{\rm sp}  L_e^{1/2}(1+\sigma_{\rm w})^{1/2}}{R\,\Gamma_w  c^{3/2} m_e}.
\label{eq:frb_frequency}
\end{equation}
where $\omega_p = (4\pi e^2 n_{\rm w}/m_e)^{1/2}$ is the plasma frequency in the rest of the CSM and we have used equation \ref{n-wind}. We can eliminate $L_e$ from this expression in terms of the observed quantities $L_{\rm FRB}$ \& $t_{\rm FRB}$ using (\ref{L-frb-shock}) \&   (\ref{t-frb-shock-maser}), and eliminate $\Gamma_{\rm sp}$ using (\ref{gamu-gamd}) to obtain:
\begin{equation}
   \nu_{\rm FRB} \equiv \frac{\omega_{\rm FRB}}{2\pi} \simeq \frac{3\,\sqrt{2} e\, \Gamma_w (L_{\rm FRB} \, t_{\rm FRB})^{1\over2} (1+\sigma_{\rm w})}{\pi  c\, m_e \varepsilon_{\rm ma}^{1/2}R^{3/2} }.
    \label{w-maser-obs2}
\end{equation} 

The cross section for induced Compton scattering (ICS) in the rest frame of the medium through which the FRB pulse travels can be written as, e.g. \cite{KumarLu-IC20}
\begin{equation}
    \sigma_{\rm ICS,w} \approx \frac{3\sigma_T \theta_{\rm s,w}^2 L_{\rm FRB,w}}{64 \pi^2 m_e \nu_{\rm FRB,w}^3 R^2},
    \label{sigma-IC3}
\end{equation}
The wind frame and observer frame variables are related as follows: $L_{\rm FRB,w}\approx L_{\rm FRB}/4\Gamma_w^2$, $\nu_{\rm FRB,w}\approx \nu_{\rm FRB}/2\Gamma_w$, $\theta_{\rm s,w} \approx \min\{\pi/2,2\Gamma_w\theta_{\rm s}\}$ and $\theta_{\rm s}$ is the opening angle of the beam. Since $4\pi R^2$ represents the surface area perpendicular to the wind direction, it is the same in primed and unprimed frames. The optical depth to ICS $\tau_{\rm ICS} = \sigma_{\rm ICS,w}n_{\rm w}\,d_{\rm w}$; where $d_{\rm w} \approx \min\left\{R/\Gamma_w, \, 2ct_{\rm FRB,w}/\theta_{\rm s,w}^2\right\}$ is the distance that a scattered photon, typically moving at an angle $\theta_{\rm s,w}$ with respect to the FRB pulse propagation direction, remains within the pulse or travels a distance where the electron density drops by a factor of $\sim 2$, whichever is smaller, and $t_{\rm FRB,w} = 2\Gamma_w t_{\rm FRB}$. Combining these factors, we find the ICS optical depth to be
\begin{equation}
    \tau_{_{\rm ICS}} \approx \frac{3\sigma_T n_{\rm w} L_{\rm FRB} c t_{\rm FRB} \Gamma_w^2 \min\bigg\{1, \frac{R\theta_{\rm s,w}^2}{2\Gamma^2_w c t_{\rm FRB}}\bigg\}}{8 \pi^2 m_e \nu_{\rm FRB}^3 R^2}
    \label{tau-ic-frb3}
\end{equation}
For maser emission from shocks in the CSM, $\theta_{\rm s,w} \approx \Gamma_w/\Gamma_{\rm sp}$. Combining this with Eqns. \ref{t-frb-shock-maser} and \ref{gamu-gamd} we find $R\theta_{\rm s,w}^2/(2\Gamma_w^2 c t_{\rm FRB}) \sim \Gamma_{\rm sf}^2/(2\Gamma_{\rm sp})^2 \sim 2(1+\sigma_{\rm w})$, and the term ``$\min\{...\}$" is equal to unity in the above equation.

Substituting for $n_{\rm w}$ using (\ref{n-wind}), and for $\nu_{\rm FRB}$ using Eq. \ref{w-maser-obs2}, we can eliminate $R$ and $L_w$ in terms of the observables $L_{\rm FRB}$, $\nu_{\rm FRB}$, and $t_{\rm FRB}$. This results in a very simple expression for the IC optical depth in the external shock model for FRBs, 
\begin{equation}
    \tau_{\rm ICS}\approx 490\, t_{\rm FRB,-3} \, \nu_{\rm FRB,9}\, \varepsilon_{\rm ma,-3}\, (1+\sigma_{\rm w})^{-2}.
 \label{tau-IC-frb7}
\end{equation}
where we have normalized $\varepsilon_{\rm ma}$ by $\sim 10^{-3}$ (e.g. \cite{Plotnikov&Sironi19} find $\varepsilon_{\rm ma}\approx 2\times 10^{-3}$). 
The induced Compton optical depth is therefore independent of the FRB luminosity, the FRB jet LF, the unshocked CSM LF, its density and the emission radius. The equation is valid at all radii, both below and above the deceleration radius of the FRB relativistic jet. The induced Compton constraint therefore cannot be evaded by tuning these quantities, and only by raising $\sigma_{\rm w}$ or accepting a high-frequency tail (see below) can the model remain viable. Specifically, since $\tau_{\rm ICS}\propto \varepsilon_{\rm ma} (1+\sigma_{\rm w})^{-2}$, reaching $\tau_{\rm ICS}\lesssim 1$ thus requires $\sigma_{\rm w}\approx 30$. However, this significantly reduces the total efficiency as discussed next.

The efficiency $\varepsilon_{\rm ma}$ was found from electron-positron PIC simulations. Assuming that the efficiency relative to the shocked electron flux is fixed when the composition is changed \footnote{ If thermal equilibrium can be reached between electrons and protons in an electron–proton plasma, the fraction of the energy flux crossing the shock front that is converted into radio waves might not be suppressed by the factor $m_p/m_e$ that one might otherwise expect. This possibility is suggested by the simulation results of \cite{Iwamoto+24}. However, if the electron–ion coupling upstream of the shock is effective, it would heat the electrons to relativistic temperatures. This, in turn, would suppress maser emission by electrons. The simulations of \cite{Iwamoto+24} were not run long enough to provide a firm answer regarding the maser efficiency in an electron–ion plasma.}, the total efficiency of the maser model is suppressed by an additional factor of $m_e/[8m_{\rm w}(1+\sigma_{\rm w})^2]$ relative to $\varepsilon_{\rm ma}$, i.e. $\varepsilon_{\rm tot}\approx 2.5\times 10^{-4} (m_e/m_{\rm w}) (1+\sigma_{\rm w})^{-2}$ (see appendix \S \ref{sec:magshock} for details) and, in turn,
\begin{equation}
\label{tau-IC-frb8}
    \tau_{\rm ICS}\approx 390 t_{\rm FRB,-3}\nu_{\rm FRB,9}\varepsilon_{\rm tot,-4}\frac{m_w}{m_e},
\end{equation}
i.e. removing the dependence on $\sigma_{\rm w}$ in this form.

\begin{figure}
\centering
\includegraphics[width=0.5\textwidth]{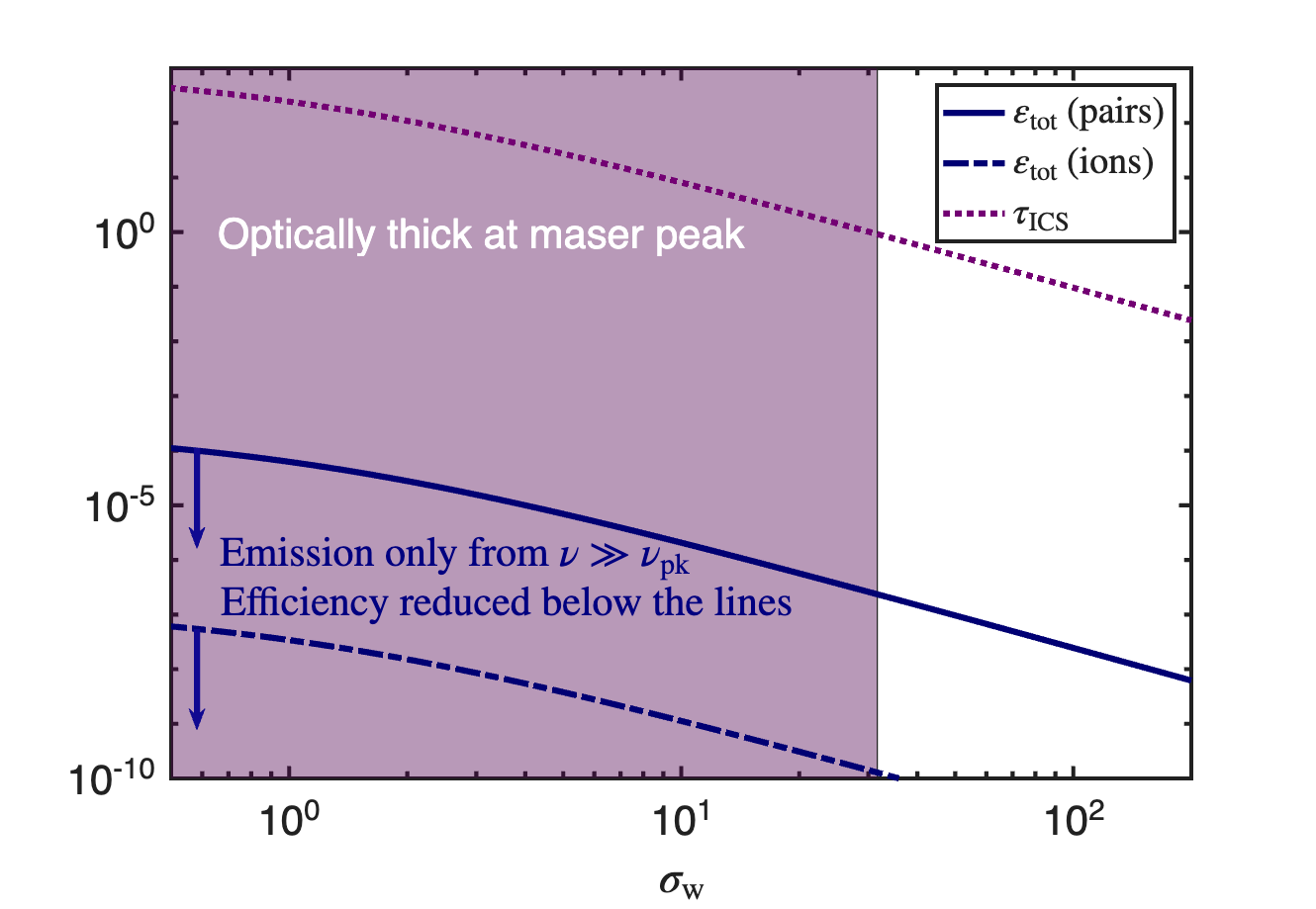}
\caption{Total radiative efficiency (blue) and optical depth to induced Compton at the peak emission frequency (purple) in the maser blastwave model. The results are shown for $t_{\rm FRB}=1$\,ms, $\nu=1$\,GHz and are independent of other FRB observables. Models involving small to moderate magnetization, $\sigma_{\rm w}\lesssim 30$, rely on intrinsically broadband spectrum of the maser emission in order to produce any observable signal. However these lead to a significant reduction of the already low efficiency, and generally results in observed broadband spectra. $\sigma_{\rm w}\gtrsim 30$ can naturally result in optically thin conditions at the maser peak, but the maximum efficiency is tiny $\epsilon_{\rm tot}\lesssim 10^{-10}$ for electron ion ($\epsilon_{\rm tot}\lesssim 2\times 10^{-7}$ for pairs).
} 
\par\vspace{1em}
\label{fig:maser}
\end{figure}

For typical FRB observables, the induced Compton optical depth is significantly larger than unity (independent of ion loading of the magnetar wind). This imposes important constraints on the synchrotron maser model in the external shock scenario. Two escape routes from this constraint appear possible, as demonstrated in Fig. \ref{fig:maser}: (i) emission at radii$\gg R$ (where $R$ is given in equation (\ref{w-maser-obs2})) via a high-frequency maser tail \citep{Metzger+19,Margalit+20}, and (ii) operation at $\sigma_{\rm w}\gtrsim 30$. Both face difficulties. (i) requires the maser to have a broadband high-frequency tail, which is poorly constrained by current PIC simulations and inconsistent with both warm-plasma simulations showing narrow $\Delta \nu/\nu\sim 0.2$ features at a frequency near that given by Eq. \ref{w-maser-obs2} \citep{Babul2020} and observations of intrinsically narrow FRB spectra. It also reduces the efficiency further relative to $\varepsilon_{\rm tot}$ discussed above. (ii) further reduces the maser efficiency as $(1+\sigma_{\rm w})^{-2}$ and shifts the typical frequency in a way that suppresses the X-ray incoherent synchrotron counterpart predicted by the model \citep{MBSM2020}.

The calculation carried out in this section implicitly assumed that the wind is not being significantly accelerated due to the momentum imparted by the FRB radiation. If the radiation accelerates the wind to $\Gamma>\Gamma_{\rm w}$, then the proper density in the wind is reduced to $n=n_{\rm w}\Gamma_{\rm w}/\Gamma$ and from Eqns. \ref{sigma-IC3},\ref{tau-ic-frb3}, $\tau_{\rm ICS,\Gamma}\approx \tau_{\rm ICS}\Gamma/\Gamma_{\rm w}$ for $\Gamma<\Gamma_{\rm sf}$ and  $\tau_{\rm ICS,\Gamma}\approx \tau_{\rm ICS}\Gamma_{\rm sf}^2/(\Gamma_{\rm w}\Gamma)$ for $\Gamma>\Gamma_{\rm sf}$. Acceleration therefore increases the optical depth until the medium outruns the shock front. Reaching $\tau_{\rm ICS}<1$ would then require $\Gamma\gtrsim \tau_{\rm ICS}\Gamma_{\rm sf}^2/\Gamma_{\rm w}\sim 10^8\Gamma_{\rm w}^{-1}$. However, even if the entire FRB energy is deposited in the upstream wind it could only account for an acceleration by an amount $\Gamma'\beta'\sim 1L_{\rm FRB,42}t_{\rm FRB,-3}R_{13}^{-3}n_{\rm w,5}^{-1}(1+\sigma_{\rm w})^{-1}$ independent of $\Gamma_{\rm w}$ and where $(1+\sigma_{\rm w})$ accounts for the inertia of the magnetic field. The values of $n_{\rm w},R$ in the above equation cannot be changed by much in the synchrotron maser model, since they are set by the requirement to reproduce the observed luminosity, emission frequency and timescale (Eqns. \ref{t-frb-shock-maser},\ref{L-frb-shock},\ref{w-maser-obs2}). Therefore radiation-pressure clearing of the upstream does not alleviate the induced-Compton constraint.

\section{Conclusions}
The majority of proposed FRB models involve the luminosity being carried at some point by a kinetic flow. We have shown here that this leads to general constraints on the mechanism powering FRBs. In particular, models in which the emission arises from near or below the light-cylinder (intermediate and magnetospheric models - including coherent-antenna mechanisms, plasma masers, and inverse-Compton-scattering scenarios), must involve continuous in-situ acceleration. Without this, their luminosity and brightness temperature fall short by orders of magnitude as compared with observed FRBs. An equivalent statement is that given the typical FRB luminosities, particles in these environments cool down on a timescale much shorter than the FRB duration, and indeed, shorter even than the FRB wave period.

Continuous acceleration can be naturally provided by an electric field, but the strength of the latter is constrained by the background magnetic field. This consideration leads to a further constraint on the emission radius, which is of the order of $10^{10}$\,cm for typical observed parameters (Eq. \ref{eq:Rmax}).
Independent of this kinetic-energy argument, the finite current that the available charges can support provides a comparable upper limit on the emission radius and applies also to conservative wave-conversion mechanisms (Eq. \ref{eq:Rcharge}).
In specific models, additional constraints can be placed based, e.g., on the emissivity per particle. A different example is that of forced reconnection near the light-cylinder. This model, while able to supply continuous acceleration, is strongly challenged to explain the observed energetics and required efficiencies of FRB sources (Eq. \ref{eq:recefficiency}).
Applying the single-pass particle-energy budget to monster shocks implies particle numbers far exceeding the magnetospheric supply and pushes the maser peak well above GHz frequencies. Very recent radiation-reaction calculations suggest that direct conversion of FMS-wave energy at the shock may relax this constraint, although nonlinear precursor feedback and wave escape remain unresolved.

Far-away models, such as synchrotron maser in a relativistic blastwave, can naturally overcome the luminosity and brightness temperature constraints mentioned above. However, the same properties of the outflow (large luminosity and density) that help it overcome these constraints, lead to an independent challenge of the synchrotron maser model in the form of the induced Compton optical depth. As we have shown here, the latter depends only on the FRB duration and frequency and on the intrinsic maser efficiency and is always $\gg 1$ (Eq. \ref{tau-IC-frb8}). The implication is that external shock maser models must rely on a significant high frequency spectral tail above the maser peak in order for any emission to come out in this scenario whatsoever. However, this then leads to a further reduced efficiency (Eq. \ref{eq:epstot}) and, more importantly, makes the common occurrence of narrow observed FRB spectra highly fine-tuned in this model.

\section*{Acknowledgments}
We thank Yuri Lyubarsky, Bing Zhang, Zorawar Wadiasingh, Alex Cooper, Kaustubh Rajwade and Jason Hessels for helpful discussions.
We acknowledge support provided by the NASA grant 80NSSC24K0770 for this work.
PB's work was also funded by a grant (no. 2024788) from the United States-Israel Binational Science Foundation (BSF), Jerusalem, Israel and by a grant (no. 1649/23) from the Israel Science Foundation.

\appendix
\restartappendixnumbering

\section{Strong Relativistic Magnetized Shocks}
\label{sec:magshock}
Consider a collision between a magnetized (with magnetization $\sigma_4\gg 1$) relativistic (with LF $\Gamma_4\gg 1$) ejecta (with comoving density $n_4$) with a magnetized relativistic wind (characterized by $\sigma_1,n_1,\Gamma_1$ accordingly).
The collision results in two shocks, a reverse shock (RS) propagating back into the ejecta and a forward shock (FS) propagating into the wind. The entire medium can be divided into four regions: unshocked wind (region 1), shocked wind (region 2), shocked ejecta (region 3) and unshocked ejecta (region 4). Regions 2 and 3 are separated by a contact discontinuity. We consider the case in which both the FS and RS are strong shocks, which holds as long as $\Gamma_{\rm RS,4}^2\gg\sigma_4$, $\Gamma_{\rm FS,1}^2\gg\sigma_1$ (where $\Gamma_{X,Y}$ denotes the LF of region $X$ relative to region $Y$). In the strong shock limit, one finds $\Gamma_{\rm RS,3}\approx \sigma_4^{1/2}, \Gamma_{\rm FS,2}\approx \sigma_1^{1/2}$.
Using the results of \cite{narayan2011}
\begin{eqnarray}
 \label{eq:app1}
  &  n_3=\frac{\Gamma_{\rm RS,4}n_4}{\Gamma_{\rm RS,3}}\quad ; \quad P_3^{\rm gas}=\frac{n_4 m_{4} c^2 \Gamma_{\rm RS,4}^2}{8\sigma_4} \quad ; \quad P_3^{\rm mag}=4\sigma_4 P_3^{\rm gas} \\
  & n_2=\frac{\Gamma_{\rm FS,1}n_1}{\Gamma_{\rm FS,2}}\quad ; \quad P_2^{\rm gas}=\frac{n_1 m_{1} c^2 \Gamma_{\rm FS,1}^2}{8\sigma_1} \quad ; \quad P_2^{\rm mag}=4\sigma_1 P_2^{\rm gas} \nonumber
\end{eqnarray}
where $m_4$ ($m_1$) is the mass of the particles dominating the rest mass of the ejecta (wind).
The total pressure in the shocked regions is:
\begin{eqnarray}
    P_3^{\rm tot}=P_3^{\rm gas}+P_3^{\rm mag}\approx P_3^{\rm mag} \quad ; \quad P_2^{\rm tot}=P_2^{\rm gas}+P_2^{\rm mag}\approx P_2^{\rm mag}
\end{eqnarray}
Equality of pressure and velocity across the contact discontinuity means that 
\begin{eqnarray}
\label{eq:pressureandGamma}
 P_3^{\rm tot}=P_2^{\rm tot}\to n_4 m_{4} \Gamma_{\rm RS,4}^2=n_1 m_{1} \Gamma_{\rm FS,1}^2 \quad ; \quad \Gamma_3=\Gamma_2.   
\end{eqnarray}
Applying a velocity transformation, we relate the shocked LFs to the LFs of the shock fronts,
\begin{eqnarray}
    &\Gamma_{\rm RS,3}\approx \frac{\Gamma_3}{2\Gamma_{\rm RS}}\to \Gamma_3=2\sigma_4^{1/2}\Gamma_{\rm RS} \quad ; \quad \Gamma_{\rm FS}\approx 2\Gamma_{\rm FS,2}\Gamma_2\to \Gamma_2\approx \frac{\Gamma_{\rm FS}}{2\sigma_1^{1/2}}  
\end{eqnarray}
and by equating the shocked LFs
\begin{equation}
\label{eq:GammaFSoverGammaRS}
   \Gamma_{\rm FS}=4\Gamma_{\rm RS} (\sigma_1 \sigma_4)^{1/2} .
\end{equation}
At the same time, we can relate the shock front LFs to the unshocked media $\Gamma_{\rm RS,4}\approx \Gamma_4/(2\Gamma_{\rm RS})$, $\Gamma_{\rm FS,1}\approx \Gamma_{\rm FS}/(2\Gamma_{1})$ and use the pressure equilibrium (left part of Eq. \ref{eq:pressureandGamma}), to find
\begin{equation}
    \Gamma_{\rm RS}\Gamma_{\rm FS}=\Gamma_1\Gamma_4 \bigg(\frac{n_4 m_4}{n_1 m_1}\bigg)^{1/2}
\end{equation}
Combining with Eq. \ref{eq:GammaFSoverGammaRS}, we arrive at expressions for the shock front LFs that depend only on unshocked flow properties
\begin{eqnarray}
    & \Gamma_{\rm FS}^2=4\Gamma_1\Gamma_4 \bigg(\frac{n_4 m_4}{n_1 m_1}\bigg)^{1/2} (\sigma_1 \sigma_4)^{1/2} \quad ; \quad  \Gamma_{\rm RS}^2=\frac{\Gamma_1\Gamma_4 \bigg(\frac{n_4 m_4}{n_1 m_1}\bigg)^{1/2}}{ 4(\sigma_1 \sigma_4)^{1/2}} \nonumber
\end{eqnarray}
It is useful to consider also the LFs of the shock fronts relative to the respective undisturbed media they are running into:
\begin{equation}
    \Gamma_{\rm FS,1}=\bigg(\frac{\Gamma_4}{\Gamma_1}\bigg)^{1/2} \bigg(\frac{n_4 m_4}{n_1 m_1}\bigg)^{1/4}(\sigma_1 \sigma_4)^{1/4} \quad ; \quad \Gamma_{\rm RS,4}=\bigg(\frac{\Gamma_4}{\Gamma_1}\bigg)^{1/2} \bigg(\frac{n_1 m_1}{n_4 m_4}\bigg)^{1/4}(\sigma_1 \sigma_4)^{1/4}
\end{equation}
The strong shock conditions $\Gamma_{\rm RS,4}^2\gg\sigma_4$, $\Gamma_{\rm FS,1}^2\gg\sigma_1$ can now be translated to a self-consistency requirement
\begin{equation}
    \bigg(\frac{\Gamma_1}{\Gamma_4}\bigg)^{2}<\frac{n_4 m_4 \sigma_4}{n_1 m_1 \sigma_1}<\bigg(\frac{\Gamma_4}{\Gamma_1}\bigg)^{2}
\end{equation}
Violation of the LHS inequality means the FS is weak and violation of the RHS inequality means the RS is weak. In particular, when the wind is non-relativistic, we see that only the latter poses a concern, and that large enough ejecta magnetization suppresses the RS. These results are consistent with the classic result by \cite{SariPiran1995} in the unmagnetized limit, which corresponds here to $\sigma_4, \sigma_1$ being $\mathcal{O}(1)$ (since we have implicitly assumed $\sigma+1\approx \sigma$ above).

In the strong shocks limit, the shocked LF is given by
\begin{equation}
    \Gamma_2=\Gamma_3\approx (\Gamma_1 \Gamma_4)^{1/2} \bigg(\frac{n_4 m_4}{n_1 m_1}\bigg)^{1/4} \bigg(\frac{\sigma_4}{\sigma_1}\bigg)^{1/4} \to \bigg(\frac{\Gamma_2}{\Gamma_1}\bigg)^4=\bigg(\frac{\Gamma_4}{\Gamma_1}\bigg)^2 \frac{n_4m_4 \sigma_4 }{n_1 m_1 \sigma_1}
\end{equation}
In particular, we can plug this into Eq.\ref{L-frb-shock}, and find (using the notation of that section, i.e. $\mbox{ej}\Leftrightarrow 4, \mbox{w}\Leftrightarrow 1$)
\begin{equation}
    L_{\rm FRB}=\varepsilon_{\rm ma} L_e \bigg(\frac{\Gamma_2}{\Gamma_1}\bigg)^4\approx  L_{\rm ej} \varepsilon_{\rm ma}\frac{m_{e}}{m_{\rm w}}(1+\sigma_{\rm w})^{-1}
\end{equation}
where $L_{\rm ej}=4\pi R^2 \sigma_{\rm ej} n_{\rm ej} m_{\rm ej}c^3\Gamma_{\rm ej}^2$ is the total luminosity of the ejecta. 

We consider next the energy of the FRB relative to that of the ejecta. The energy density in the shocked wind frame (region 2) is $u_2\approx n_{\rm 1}(1+\sigma_{\rm 1})m_{\rm 1} c^2\bigg(\frac{\Gamma_2}{\Gamma_{\rm 1}}\bigg)^2$. In the central engine frame, the energy density is enhanced by a factor of $\Gamma_2^2$. The volume of the shocked region in the same frame is $V\sim 4\pi R^3/\Gamma_2^2$. Put together, we find that the total energy of the shocked plasma in region 2 is
\begin{equation}
    E_2\approx 4\pi R^3 n_1 m_1 (1+\sigma_1)c^2\bigg(\frac{\Gamma_2}{\Gamma_{\rm 1}}\bigg)^2 
\end{equation}
At the deceleration radius this energy is comparable to the ejecta energy, $E_2=E_{\rm ej}$. Re-expressing in terms of $L_{\rm FRB}$ and $t_{\rm FRB}$ and adjusting the notation as above, we  get
\begin{equation}
    E_{\rm FRB}=L_{\rm FRB}t_{\rm FRB}=E_{\rm ej}\frac{1}{8}\varepsilon_{\rm ma}\frac{m_{e}}{m_{\rm w}}(1+\sigma_{\rm w})^{-2}.
\end{equation}

Therefore, we find that the total efficiency of synchrotron maser (in the sense of converting energy being produced by the source to maser energy being radiated away), is 
\begin{equation}
\label{eq:epstot}
    \varepsilon_{\rm tot}=\frac{1}{8}\varepsilon_{\rm ma} \frac{m_e}{m_{\rm w}}(1+\sigma_{\rm w})^{-2}. 
\end{equation}

This energy-based efficiency is smaller by a factor of $8(1+\sigma_{\rm w})\gg 1$ as compared to the luminosity-based efficiency. This is a reflection of the fact that the duration of the FRB is smaller by the same factor as compared with the duration of the ejecta.

The efficiency estimate can be understood physically as arising from the fact that the magnetization parameters on the two sides of a shock front are equal up to a factor of order unity (see eq. \ref{eq:app1}). Thus, the fraction of energy carried by electrons in region 2 is $m_e/(m_{\rm w} \sigma_{\rm w})$ relative to the total in that region. In addition, beyond the deceleration radius, more than half of the energy of the magnetar ejecta resides in the shocked CSM, i.e. region 2. 
It follows, therefore, that a fraction $\epsilon_{\rm ma}$ of the electron luminosity in region 2 is converted to radio waves. Since the observed duration is compressed by a factor of $(1+\sigma_{\rm w})$, we see that overall a fraction $\sim \epsilon_{\rm ma} m_e/[m_{\rm w} (1+\sigma_{\rm w})^2]$ of the ejecta's energy is converted into radio waves.

\section{Conservative conversion of low-frequency waves to GHz radiation}
\label{sec:conserve}
\subsection{Weakly nonlinear conversion}
Consider an electromagnetic pump whose characteristic frequency $\omega_0$ is much smaller than the observed FRB frequency, $\omega_{\rm FRB}$. A possible way to avoid the kinetic-energy bottleneck discussed in \S \ref{sec:applicable} is conservative wave-wave conversion, in which a material current mediates transfer of energy between electromagnetic modes while acquiring little net energy itself. In the weakly nonlinear regime, such interactions are constrained by the usual resonance conditions. For a three-wave process, $\omega_3=\omega_1+\omega_2$ and ${\bf k}_3={\bf k}_1+{\bf k}_2$ with finite interaction time and length allowing sufficiently small mismatches: $|\Delta\omega|t_{\rm int}\lesssim 1$, $|\Delta k|\ell_{\rm int}\lesssim 1$. The corresponding Manley-Rowe relations \citep{1956PIRE...44..904M} describe reversible exchange of wave action and allow energy transfer between modes without significant energy passing via the particles.
Direct conversion from $\omega_0$ to $\omega_{\rm FRB}\gg\omega_0$, however, is a very high-order process. If $\mathcal{N}=\omega_{\rm FRB}/\omega_0\gg 1$, direct production of the $\mathcal{N}$-th harmonic arises only at high order in a perturbative expansion and is therefore strongly suppressed when the nonlinear perturbation amplitude is small. Alternatively, one may consider a sequence of low-order interactions through intermediate modes. Even in the optimistic limit in which the characteristic frequency doubles at every step, the minimum number of successive frequency-doubling stages (or generations) required is $N_{\rm st}=\log_2(\omega_{\rm FRB}/\omega_0)\approx 20$ for an expected $\omega_{\rm FRB}/\omega_0\sim 10^6$. Such a sequence is limited by the requirement of appropriate resonant daughter modes and phase matching over a broad range of frequencies.

\subsection{Strong nonlinear cascades and spectral selection}
Strongly nonlinear evolution provides a more general possibility. When the nonlinear interaction rate $\Gamma_{\rm nl}$ becomes sufficiently large, resonances are broadened and many modes may interact simultaneously, so that the description in terms of a sequence of isolated three-wave processes is no longer appropriate. Energy can then cascade across a broad range of frequencies without the perturbative suppression associated with direct high-order harmonic generation. We therefore cannot exclude such a conservative route to GHz radiation on general grounds.
There is nevertheless a tension between rapid broadband transfer and the relatively narrow instantaneous spectra observed in many FRBs. Nonlinear interactions broaden a mode by an amount of order $\Delta \omega=\Gamma_{\rm nl}$ where $\Gamma_{\rm nl}$ is the non-linear transfer rate. If the terminal nonlinear interaction itself sets the observed fractional FRB spectral bandwidth, $\delta=\Delta \nu/\nu$, then the requirement becomes $\Gamma_{\rm nl}/\omega_{\rm FRB}<\delta$. A strongly nonlinear cascade capable of rapidly transporting energy over many decades in frequency is therefore not by itself guaranteed to produce a narrow GHz feature. A successful model generally requires the cascade to terminate in, or couple efficiently to, a frequency-selective process such as a plasma resonance, cutoff, or other characteristic scale near the observed frequency. The existence of such mechanisms is model dependent, and we do not attempt to exclude them here.

\end{document}